\documentclass[fleqn,usenatbib]{mnras}

\usepackage{newtxtext,newtxmath}

\usepackage[T1]{fontenc}
\usepackage{ae,aecompl}
\usepackage[dvipsnames]{xcolor}
\usepackage{comment}

\usepackage{graphicx}	
\usepackage{amsmath}	
\usepackage{amssymb}	
\usepackage{multicol}	

\newcommand{\beq}{\begin{equation}}
\newcommand{\eneq}{\end{equation}}
\newcommand{\becase}{\begin{cases}}
\newcommand{\encase}{\end{cases}}
\newcommand{\bet}{\begin{table}}
\newcommand{\ent}{\end{table}}

\title[Formation of the Galilean Satellites]{Satellites Form Fast \& Late: a Population Synthesis for the Galilean Moons}

\author[Cilibrasi et al.]{
M. Cilibrasi$^{1,2,3}$\thanks{E-mail: marco.cilibrasi@alumni.sns.it},
J. Szul\'agyi$^{3,4}$,
L. Mayer$^{3}$,
J. Dr\c{a}\.{z}kowska$^{3,5}$,
Y. Miguel$^{6,7}$,
P. Inderbitzi$^{3}$
\\
$^{1}$Scuola Normale Superiore, Piazza dei Cavalieri 7, I-56126 Pisa, Italy\\
$^{2}$Dipartimento di Fisica "Enrico Fermi", Universit\`a di Pisa, Largo Pontecorvo 3, I-56127 Pisa, Italy\\
$^{3}$Centre for Theoretical Astrophysics and Cosmology, Institute for Computational Science, University of Z\"urich,\\ Winterthurerstrasse 190, CH-8057 Z\"urich, Switzerland\\
$^{4}$ETH Z\"urich, Institute for Particle Physics and Astrophysics, Wolfgang-Pauli-Strasse 27, CH-8093, Z\"urich, Switzerland\\
$^{5}$University Observatory, Faculty of Physics, Ludwig-Maximilians-Universit\"{a}t M\"{u}nchen, Scheinerstr. 1, 81679 Munich, Germany\\
$^{6}$Observatoire de la C\^ote d'Azur, Boulevard de l'Observatoire, CS 34229 F-06304, Nice Cedex 4, France\\
$^{7}$Leiden Observatory, University of Leiden, Niels Bohrweg 2, NL-2333CA Leiden, The Netherlands\\
}

\date{Accepted 2018 August 6. Received 2018 August 6; in original form 2018 January 10}

\pubyear{2018}

\begin{document}
\label{firstpage}
\pagerange{\pageref{firstpage}--\pageref{lastpage}}
\maketitle

\begin{abstract}
The satellites of Jupiter are thought to form in a circumplanetary disc. Here we address their formation and orbital evolution with a population synthesis approach, by varying the dust-to-gas ratio, the disc dispersal timescale and the dust refilling timescale. The circumplanetary disc initial conditions (density and temperature) are directly drawn from the results of 3D radiative hydrodynamical simulations. The disc evolution is taken into account within the population synthesis. The satellitesimals were assumed to grow via streaming instability. 
We find that the moons form fast, often within $10^4$ years, due to the short orbital timescales in the circumplanetary disc. They form in sequence, and many are lost into the planet due to fast type I migration, polluting Jupiter's envelope with typically 15 Earth-masses of metals. The last generation of moons can form very late in the evolution of the giant planet, when the disc has already lost more than the 99\% of its mass. The late circumplanetary disc is cold enough to sustain water ice, hence not surprisingly the 85\% of the moon population has icy composition. The distribution of the satellite-masses is peaking slightly above Galilean masses, up until a few Earth-masses, in a regime which is observable with the current instrumentation around Jupiter-analog exoplanets orbiting sufficiently close to their host stars. We also find that systems with Galilean-like masses occur in $20\%$ of the cases and they are more likely when discs have long dispersion timescales and high dust-to-gas ratios.

\end{abstract}




\section{Introduction}\label{introduction}

In the last few years theories about our Solar System formation took a step forward thanks to a more precise comprehension of giant planet formation and evolution within protoplanetary discs. Today the two main models in this field are the Gravitational Instability scenario, or GI \citep{Boss97,Durisen07}, when a self-gravitating gaseous clump directly collapses into a giant planet, and the Core Accretion model, or CA \citep{Pollack96}, that occurs when collisions and coagulation of dust particles form a solid planetary embryo, massive enough to accrete and maintain a gaseous envelope. Both of these theories predict the presence of circumplanetary discs (CPDs) made of gas and dust rotating around the forming planet in the last stage of formation \citep{Alibert05,AB09,Ward10,Szulagyi17a,Szulagyi17b}. 
Even though these discs are similar to protoplanetary discs (PPDs) around young stars, there are significant differences among them. The most important one is that the CPDs are continuously fed by a vertical influx of gas and well coupled dust from the protoplanetary disc upper layers, due to gas accretion onto the central giant planet \citep{Tanigawa12,Szulagyi14}.

Due to the fact that regular satellites (including the moons of Jupiter) are commonly thought to form in CPDs, the understanding of the properties of these discs is crucial to address satellite formation. With no observational constraints about them, so far we have to rely on hydrodynamic simulations to study the initial CPD that have formed the Galilean satellites (e.g. \citealt{AB09,Gressel13,Szulagyi17b}). The properties of Jupiter's four biggest moons, however, provide some constraints about the features of this disc. Voyager and Galileo missions revealed that Io is rocky, while the outer three moons contain significant amount of water ice  \citep{Showman99}. The accretion of icy satelletesimals is only possible in a CPD which has a bulk temperature below the water freezing point, $\sim$180 K. However, hydrodynamic simulations of CPDs found the temperature to be significantly higher than that, often peaking at several thousands of Kelvins (e.g. \citealt{AB09,Szulagyi16}). The study of \citet{Szulagyi17} showed that even accounting for the cooling of the planet (due to radiating away its formation heat), the Jupiter surface temperature had to be significantly lower than 1000 K, when the Galilean satellites have formed, otherwise the CPD cannot form icy satellites. This indicated that the moons had to form very late in the planet- \& disc-evolution, when  Jupiter has significantly cooled off and its CPD was dissipating (moving towards the optically thin, and hence cold regime). 

Regarding the mass of the CPD, we know that the total mass of the Galilean satellites is $\sim 2 \times 10^{-4} M_{\rm{planet}}$ ($M_p$ hereafter), same as in the case of Saturn \citep{Canup06}. Because this value considers only solids, with a standard dust-to-gas ratio of 0.01 one gets a CPD mass of $\sim 2 \times 10^{-2} M_{p}$. However, as the Canup \& Ward works have pointed out \citep{Canup02,Canup06,Canup09}, this is the integrated CPD mass, i.e. at a snapshot of time the CPD can be much lighter than this while still producing Galilean mass satellites over the years (\textit{gas-starved disc} model). Due to the continuous feeding from the protoplanetary disc, throughout the lifetime of the CPD, even orders of magnitude more material could have been processed through the CPD. The mass of the disc has been certainly enough to make several generations of Galilean-mass moons, and several of them could have been lost into the planet through migration, opening the idea of \textit{sequential satellite-formation} \citep{Canup02}.

There have been several different approaches to study satellite formation, starting from works that studied conditions of the CPD during satellite formation and constraints on this disc based on the properties of the Galilean moons \citep{Canup09,Canup02,Estrada09}. Recently, \citet{Fujii17} numerically solved a 1D-model of circumplanetary disc long term evolution and the migration of satellites in it. They found that the moonlets are often captured in resonances, which could explain the formation of the first three resonant satellites. A population synthesis work made by \citet{Sasaki10} modeled the initial circumplanetary disc density profile solving a 1D equation for its viscous evolution \citep{Pringle81} with an inner cavity between the planet and disc. They included satellite accretion with gravitational focusing and the type I migration timescale using the formula from \citealt{Tanaka02}. Building a semi-analytical model and performing a population synthesis varying the location of the initial seeds, the $\alpha$ viscosity and the dispersion time of the disc, they found that in 70\% of their runs they had 4 or 5 satellites, often locked in a resonant configuration thanks to the inner cavity of the disc. They varied the initial circumplanetary disc profiles and used quite different models than what recent hydrodynamic models on the CPD predict (e.g. \citealt{AB09}, \citealt{Tanigawa12}, \citealt{Szulagyi14}, \citealt{Szulagyi17}). Same is true for the \citet{Miguel16}, which used the Minimum Mass SubNebula (MMSN) as an initial CPD profile \citep{Mosqueira03}. They studied the evolution of about 20 satellite-seeds, with initial positions randomly chosen in the disc, together with the gas density of the disc (but without the temperature evolution in their case), considering also the dust depletion caused by the accretion of dust itself onto protosatellites. Different runs have been made with different disc parameters, such as the dust-to-gas ratio of the disc, its dispersion timescale and the initial mass of satellitesimals, using then a population synthesis approach to analyse the outcomes.

Other different approaches to satellite formation around gas giants are, for instances in \citet{Heller15a,Heller15b} and in \citet{Moraes17}. In the first two works the authors built a semi-analytical model for the CPD structure and evolution, considering different heating processes (viscous heating, accretion onto the CPD, radiation from the hot planet, and radiation from the star) in order to study how the ice line changes position over time during the evolution of the whole system. They found that in order to reproduce the actual compositions of the Galilean satellites, the moons themselves have to form late in Jupiter evolution, when the disc is sufficiently cold, that agrees with the sequential formation process. In the latter case the authors performed N-body simulations with satellitesimals in a minimum mass subnebula circumplanetary disc, considering various eccentricities and inclinations of their orbits. In their simulations they found that satellites should never open a gap in the CPD because they are not massive enough and that initial eccentricities and inclinations do not affect results because of the gas damping. 

Because previous works have used CPD profiles that were derived from the current composition and location of the Galilean satellites without taking into account their migration and the possibility for several lost satellites system, here we present a population synthesis \citep{Benz14} on CPD profiles that are consistent with recent radiative hydrodynamical simulations on the circum-Jovian disc. We also take into account the thermal evolution of the disc, and the continuous feeding of gas and dust from the vertical influx from the protoplanetary disc (e.g. \citealt{Tanigawa12, Szulagyi14, FC16}). Moreover, we use a dust-coagulation and evolution code to calculate the initial dust density profile corresponding to the gas hydrodynamics of \citet{Szulagyi17}. We assumed that the initial seeds were formed via streaming instability \citep[e.g.][]{Youdin05}, and we placed these moonlets at the location where the conditions for streaming instability are satisfied (e.g. the local dust-to-gas ratio is higher than unity).

\section{Methods}\label{methods}

\subsection{Hydrodynamic simulation}
\label{sec:hydro}
For the circumplanetary  disc density and temperature profiles we used a simulation from \citet{Szulagyi17}. Among the various models in that paper considering different planetary temperatures, we used here one of the coldest (most evolved) state with planetary temperature of 2000 K. This is because the satellites of Jupiter are icy, they had to form in a cold circumplanetary  disc, when the planet has cooled off efficiently \citep{Szulagyi17}. This is only true in the very late stage of circumplanetary disc evolution, close to the time when the circumstellar disc has dissipated away. 

Our hydrodynamic simulation was performed with the JUPITER hydrodynamic code \citep{Borro06,Szulagyi16} developed by F. Masset \& J. Szul\'agyi. This code is three dimensional, grid-based, uses the finite-volume method and solves the Euler equations, the total energy equation and the radiative transfer with flux limited diffusion approximation, according to the two-temperature approach \citep[e.g.][]{Kley89,Commercon11}. The simulation contained a circumstellar  disc between 2.08 AU till 12.40 AU (sampled in 215 cells radially), with an initial opening angle of 7.4 degrees (from the midplane to the  disc surface, using 20 cells). The coordinate system in the simulation was spherical, centreed on the Sun-like star and co-rotating with the planet. The initial surface density was a power-law function with  $2222 \rm{kg m^{-2}}$ at the planet's location at 5.2 AU and an exponent of -0.5. The planet was a Jupiter analog, which reached its final mass through 30 orbits. The circumstellar  disc azimuthally ranged over 2$\pi$ sampled into 680 cells. To have sufficient resolution on the circumplanetary  disc developed around the gas-giant, we placed 6 nested meshes around the planet, each doubling the resolution in each spatial direction. Therefore, on the highest resolution mesh the sampling was $\sim 80\%$ of Jupiter-diameter ($\sim112000$ km) for a cell-diagonal. For the boundaries and resolution of each refined level, we used the same as Table 1 in \citet{Szulagyi16}. Because the resolution is sub-planet resolution, at the planet location we fixed the temperature to 2000\,K (thereafter referred as planet temperature) within 3 $\rm{R_{Jupiter}}$, corresponding an evolved, late stage of the circumstellar disc and planet system, roughly around 1-2 Myrs \citep{Mordasini17}. 

The equation of state in the simulation was ideal gas -- $P=(\gamma-1)\mathrm{E_{int}}$ --  which connects the internal energy ($\mathrm{E_{int}}$) with the pressure (P) through the adiabatic exponent: $\gamma=1.43$. For the viscosity, we solve the viscous stress tensor to set a constant, kinematic (physical) viscosity, that equals to 0.004 $\alpha$-viscosity at the planet location. Due to the radiative module and the energy equation, the gas can heat up through viscous heating, adiabatic compression and cool through radiation and adiabatic expansion. The opacity table used in the code was of \citet{BL94} that contains both the gas and dust Rosseland-mean opacities. Therefore, even though there is no dust component explicitly included into the simulations, the dust contribution to the temperature is taken into account through the dust-to-gas ratio, that was chosen to be 0.01, i.e. equal to the interstellar medium value \citep{Boulanger00}. The mean-molecular weight was set to 2.3, which corresponds to solar composition. The rest of the parameters and process of the simulation can be found in \citet{Szulagyi17} and \citet{Szulagyi16}.

\subsection{Population synthesis}

Our semi-analytical model essentially consists of a circumplanetary disc in which protosatellites can migrate, accrete mass and be lost into the central planet. In the meantime, while the disc density and temperature evolve in time, it creates newer and newer protosatellites. The units in our population synthesis are the following: $R_p$ as planet radius, $M_p$ (planetary mass), time in years and temperature in Kelvin. When referring to the specific case of Jupiter, $M_J$ is used as the unit for masses.

\subsubsection{Disc structure}\label{disc_structure}

\begin{figure} 
\includegraphics[width=\columnwidth]{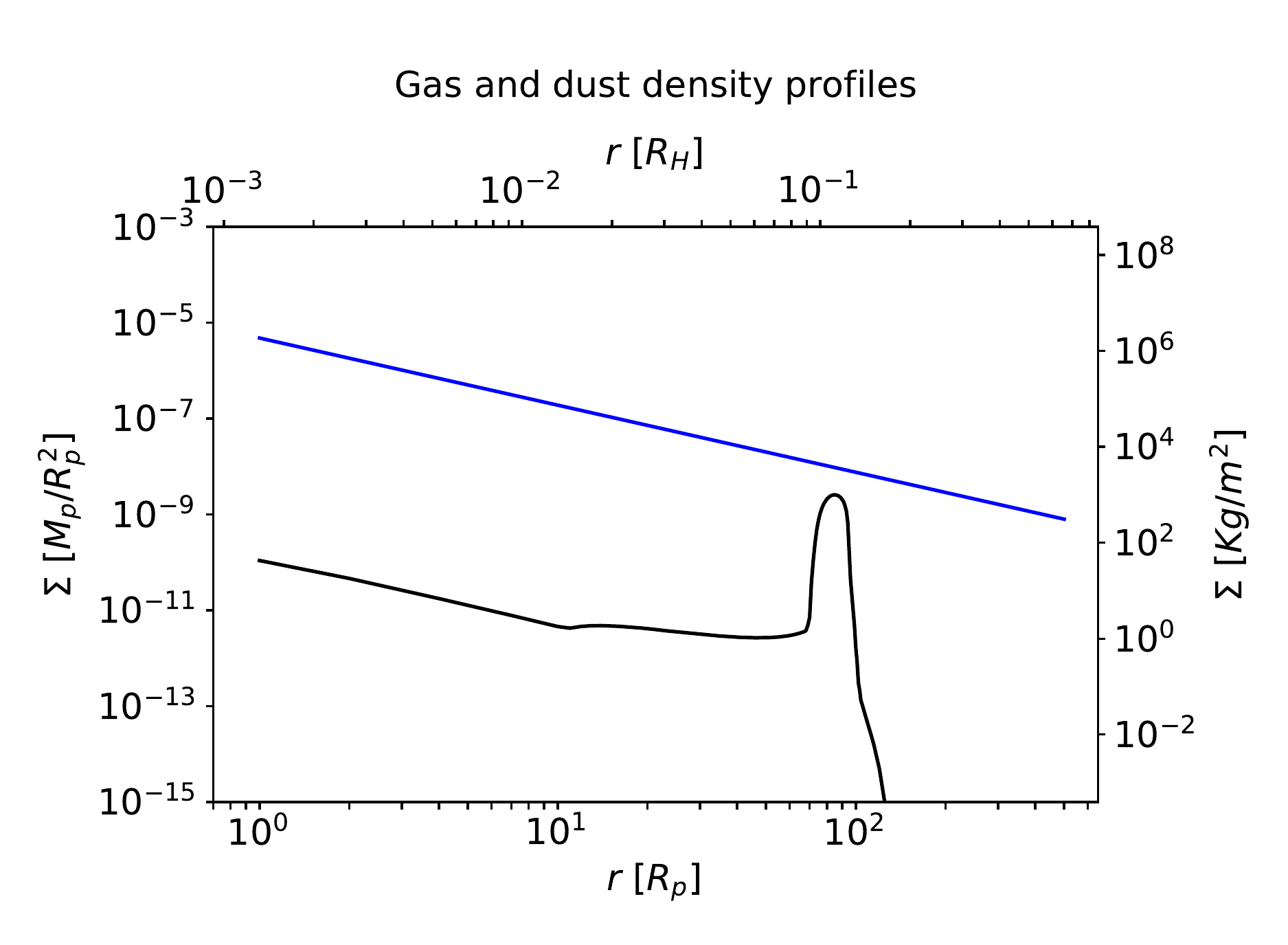}
\caption{Gas (blue) and dust (black) density profiles of the circumplanetary  disc at the beginning of the population synthesis. The dust-to-gas ratio here was chosen to be $0.08$, but this parameter is varied in the population synthesis.}
\label{disc.density}
\end{figure}

\begin{figure} 
\includegraphics[width=\columnwidth]{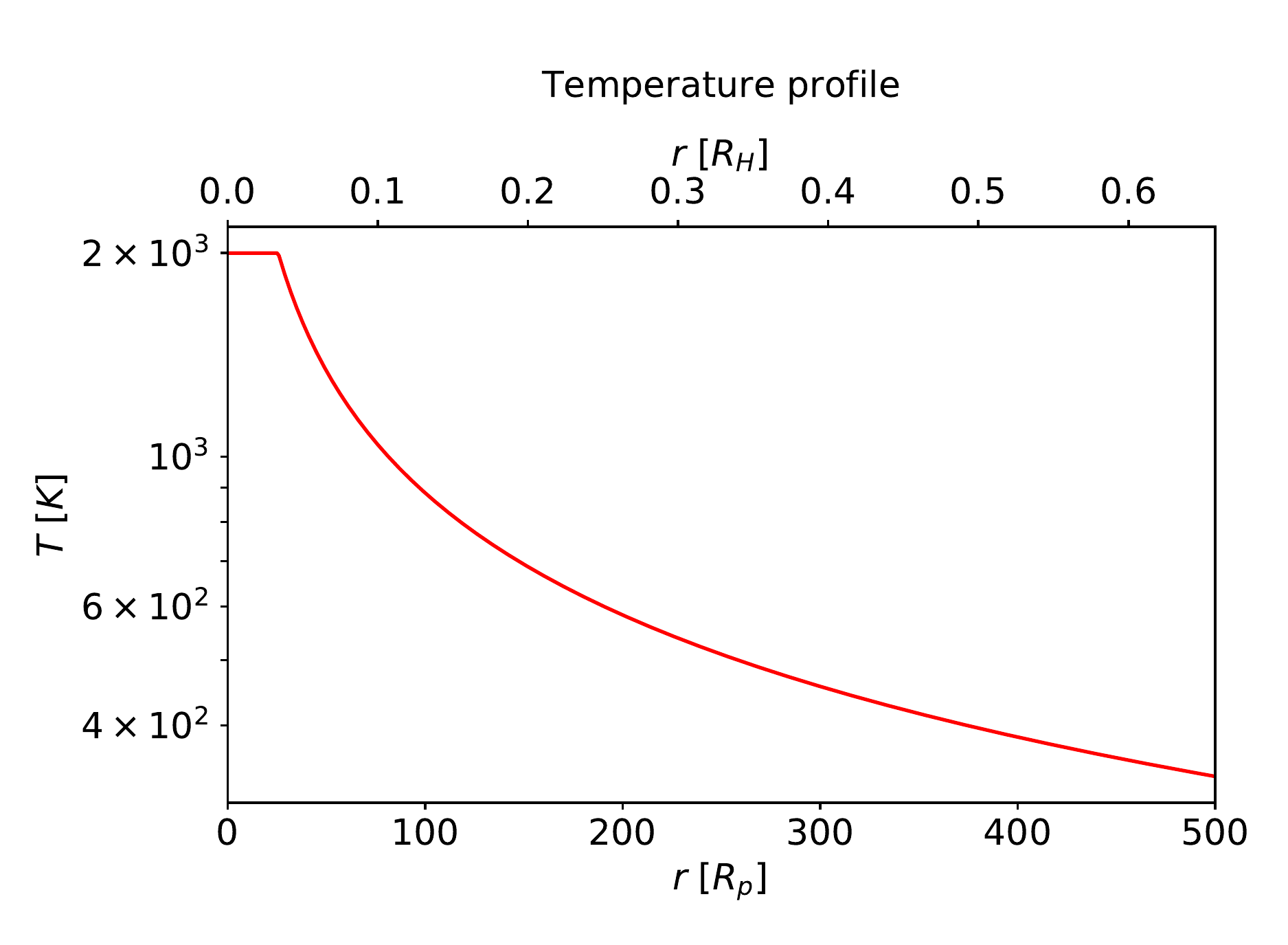}
\caption{Temperature profile of the  disc at the beginning of the population synthesis. $T_{max}=2000 K$ corresponds to the planet temperature, while $T_{min}=130K$ is the background temperature at Jupiter's location (5.2 AU from the Sun).}
\label{disc.temperature}
\end{figure}

In the model, the CPD is simply defined by its surface density (both solid and gas) profiles, temperature profile and other quantities, such as $\alpha$ for viscosity, $\gamma$ for heat capacity ratio and $C_V$ for heat capacity at constant volume. All other quantities in the  disc, such as the angular velocity of the gas, the height of the  disc, the speed of sound, etc., are computed starting from temperature and density values and using the common 1D model for discs \citep{Pringle81}. The  disc ranges between $1R_p$ and $500R_p$, according to the hydrodynamical simulation, and it is divided in 500 cells. In our model we do not consider a cavity between the planet's surface and the disc, because the magnetic field of the planet and the ionization of the disc are probably not strong enough to produce such a cavity (see also in Section \ref{discussion}). The disc initial temperature and gas density profiles are power-law fits to the results of a radiative hydrodynamical simulation of \citet{Szulagyi17} with planet temperature of 2000 K (i.e. a late time in the evolution of the forming planet \& its disc, corresponding to roughly 1-2 Myrs of PPD age), described in Section \ref{sec:hydro}. The power-laws are the followings (Figure \ref{disc.density}, Figure \ref{disc.temperature}):

\begin{equation}
\Sigma_{gas}(r)\simeq 4.8 \cdot 10^{-6} \left(\frac{r}{R_p} \right)^{-1.4} \left[\frac{M_p}{R_p^2} \right]
\end{equation}
\begin{equation}
T(r)=\begin{cases}
1.4\cdot10^4 \left(\frac{r}{R_p} \right)^{-0.6}[K] &T_{min}<T<T_{max}\\
T_{min} &T\le T_{min}\\
T_{max} &T\ge T_{max}
\end{cases}
\end{equation}
with $T_{min}=130\,K$, that is the background temperature in the PPD at Jupiter's location like e.g. in \citet{Miguel16}, and $T_{max}=2000\,K$, that is the planet temperature in the simulations. The total disc mass is $M_0\simeq 2 \times 10^{-3}M_p$, always accordingly to the 3D hydrodynamic simulation. Other parameters are chosen to be consistent with the hydrodynamic simulation, therefore the viscosity is $\alpha=0.004$ and it is considered constant in all the model, the adiabatic index is $\gamma=7/5$ (i.e. molecular hydrogen) and the heat capacity ($C_V$) equals to $10.16KJ/(KgK)$, again because of consistency with the hydro simulation.

Because the hydrodynamical simulation only gives gas density profile, we used the dust density profile of \citet{Drazkowska18}. In this work, the dust evolution code was run in a post-processing step, using the same CPD properties as we use in this paper. The surface density of dust was obtained by solving the advection-diffusion equation and applying a simplified algorithm for dust growth based on the work of \citet{Birnstiel12}. Satellitesimal formation via streaming instability was included using a method analogical as in \citet{Drazkowska16}. The dust infall onto the CPD was included assuming that the dust and gas are well mixed, so the dust infall profile is the same as the gas infall profile. The dust profile that we used in the population synthesis is an equilibrium profile resulting from the balance between dust infall, advection, diffusion, and satellitesimal formation. The dust evolution timescale in the CPD is very short, on the order of 1 year. Therefore, even though at the very beginning of dust evolution simulation dust density gets larger, the evolution finds an equilibrium after a few dozens of orbits of the CPD. The limitation of dust code is that the gas density profile (coming from the same hydro simulation, shown in Figure \ref{disc.density}) is kept constant, what is however justified by the short timescale of dust evolution. We tested than an analogical dust profile is recovered in a case when the gas surface density and infall are reduced by 50\%.

As the figure shows, there is a peak in the dust density  profile at around $85 R_J$. This dust trap is the consequence of the position where the radial velocity of gas changes sign in the hydro simulation, i.e. the gas is bringing small dust particles from the inner and outer disc to this location. The temperature of the dust was assumed to be the same as the gas temperature, assuming perfect thermal equilibrium.

Given that the dust-to-gas ratio of the CPD is not known, we kept it as a free parameter in the population synthesis. Therefore the dust density profiles were multiplied by a scalar in each individual run of the population synthesis. This is not exactly accurate since simulations do not show a simple linear scaling, but we checked that this assumption does not change results significantly. In Figure \ref{disc.density} the dust-to-gas ratio at the equilibrium is $0.08$. In this work we always refer to this final ratio, however \citet{Drazkowska18} found that this final and equilibrium value is about $5.8$ times the initial dust-to-gas ratio of the simulated disc.

In our model, we do not consider the jump of the solid density beyond the ice line, that could happen due to the condensation of water (see for example \citealt{Mosqueira03,Miguel16,Drazkowska17}). In fact, the code of \citet{Drazkowska18} found that the dust dynamics is dominated by the gas flow and thus, even after the ice line enters the disc because of cooling, the modification to the dust surface density (in Figure \ref{disc.density}) due to the existence of solid ice is negligible, it only affects the composition of the dust profile in terms of volatiles. Since we are not interested in the exact composition of the dust (see Section \ref{T_form} for more details) we simply neglected the possibility to have a dust density-jump at the ice line. 

\subsubsection{Disc evolution}\label{disc_evolution}

We adopted a self-similar solution for disc evolution. It is known that the disc is fed by the vertical influx from the protoplanetary disc \citep{Tanigawa12, Szulagyi14} that should decrease exponentially with time as the PPD dissipates \citep{Ida08}: $\dot{M}_{in}=\dot{M}_{in,0}e^{-t/t_{\rm{disp}}}$, where $t_{\rm{disp}}$ is the characteristic dispersion time of PPD and $\dot{M}_{in,0}\simeq 2 \times 10^{-6} M_p/yr$ in our case, in agreement with numerical simulations in Section \ref{sec:hydro}. The mass loss is assumed to be proportional to the mass of the disc itself: $\dot{M}_{out}=\dot{M}_{out,0}\frac{M}{M_0}$. We also assume that the CPD is initially at the equilibrium, i.e. $\dot{M}_{in,0}=\dot{M}_{out,0}=\dot{M}_0$ (accretion rates at time zero). Solving the  equation 
\beq
\frac{dM}{dt}=\dot{M}_{in}-\dot{M}_{out}
\eneq
one can find that if $t_{\rm{disp}}\gg M_0/\dot{M}_0$, and it is always the case in our model (see the values for $t_{\rm{disp}}$ in the next paragraph), the CPD density decreases exponentially with $t/t_{\rm{disp}}$ keeping the equilibrium configuration, following in practice the decrease of $\dot{M}_{in}(t)$. Therefore, in our population synthesis, the disc density evolution for both the gas profile \textit{'g'} and the solid (dust) profile \textit{'s'} is given by:
\begin{equation}
\begin{cases}
\Sigma_g=\Sigma_{g,0}e^{-t/t_{\rm{disp}}}\\
\Sigma_s=\Sigma_{s,0}e^{-t/t_{\rm{disp}}}-A\\
\end{cases}
\end{equation}
where $t_{\rm{disp}}$ is the dispersion time of the CPD (that is equal to the dispersion time of the PPD), $\Sigma_{g,0}(r)$ and $\Sigma_{s,0}(r)$ are the initial density profiles, while $A$ is the dust accreted by the protosatellites and then regenerated by the refilling mechanism, as it will be explained in Section \ref{form_and_evo}. This means that, except the term A, both the gas and dust profile will decrease following a self-similar solution, keeping the shape shown in Figure \ref{disc.density}. It is important here to keep in mind that this solution from \citet{Ida08} is valid when $\alpha$ parameter for viscosity is constant, like in our case.

The disc dispersion timescale and the total disc lifetime are not the same thing but they are not independent from each other as well, hence we also linked them in our calculation. Recent observations showed that disc lifetimes distribute exponentially between $1 Myr$ and $10 Myr$ with a characteristic age of $2.3 Myr$ \citep{Fedele10, Mamajek09}. These surveys have an accreation rate sensitivity limit till $> 10^{-11} M_{\odot} yr^{-1}$, however, on average, young T Tauri stars with a protoplanetary disc show an accretion rate of $\sim 10^{-7} M_{\odot} yr^{-1}$ (e.g.  \citealt{Ercolano14}). Considering these limits, and considering the exponential evolution of disc density (and mass), the disc lifetime will be:
\beq
t_{\rm{lifetime}}=-t_{\rm{disp}}\,ln\left(\frac{10^{-11} M_{\odot} yr^{-1}}{10^{-7} M_{\odot} yr^{-1}}\right) \simeq 10t_{\rm{disp}}
\eneq
where the dispersion timescales are distributed exponentially between $0.1 Myr$ and $1.0 Myr$, with a mean of $0.23 Myr$.

The temperature evolution was calculated also with an exponential decrease to be consistent with the density evolution:
\begin{equation}
T=T_{min}+(T_0-T_{min})e^{-t/t_{\rm{cool}}}
\end{equation}
where $t_{\rm{cool}}$ is computed with the radiative cooling formula of \citet{Wilkins12}:
\begin{equation}
\dot{T}\propto\dot{U}=-\sigma\frac{T^4-T_{min}^4}{\Sigma_g(\tau+\tau^{-1})}
\end{equation}
The optical depth ($\tau$) can be estimated as $\tau=\int \rho \kappa dh \simeq \kappa \Sigma_g$, where $\kappa(\Sigma,T)$ is the opacity computed with tables in \citet{Zhu09}, that are based on a ISM mean dust-to-gas ratio of 1\%, that is kept constant in this temperature evolution calculation, i.e. we do not consider depletion because of satellite accretion. This choice was made because the initial dust-to-gas ratio in the CPD is highly unknown, furthermore, the dust density is highly variable during the evolution of a system. We also tried to add viscous heating in the cooling model ($C_V\dot{T}=\alpha c_s^2 \Omega$, but since it did not change the result significantly, we omitted it in the final runs to save computational time.)

\begin{figure}
\centering
\includegraphics[width=\columnwidth]{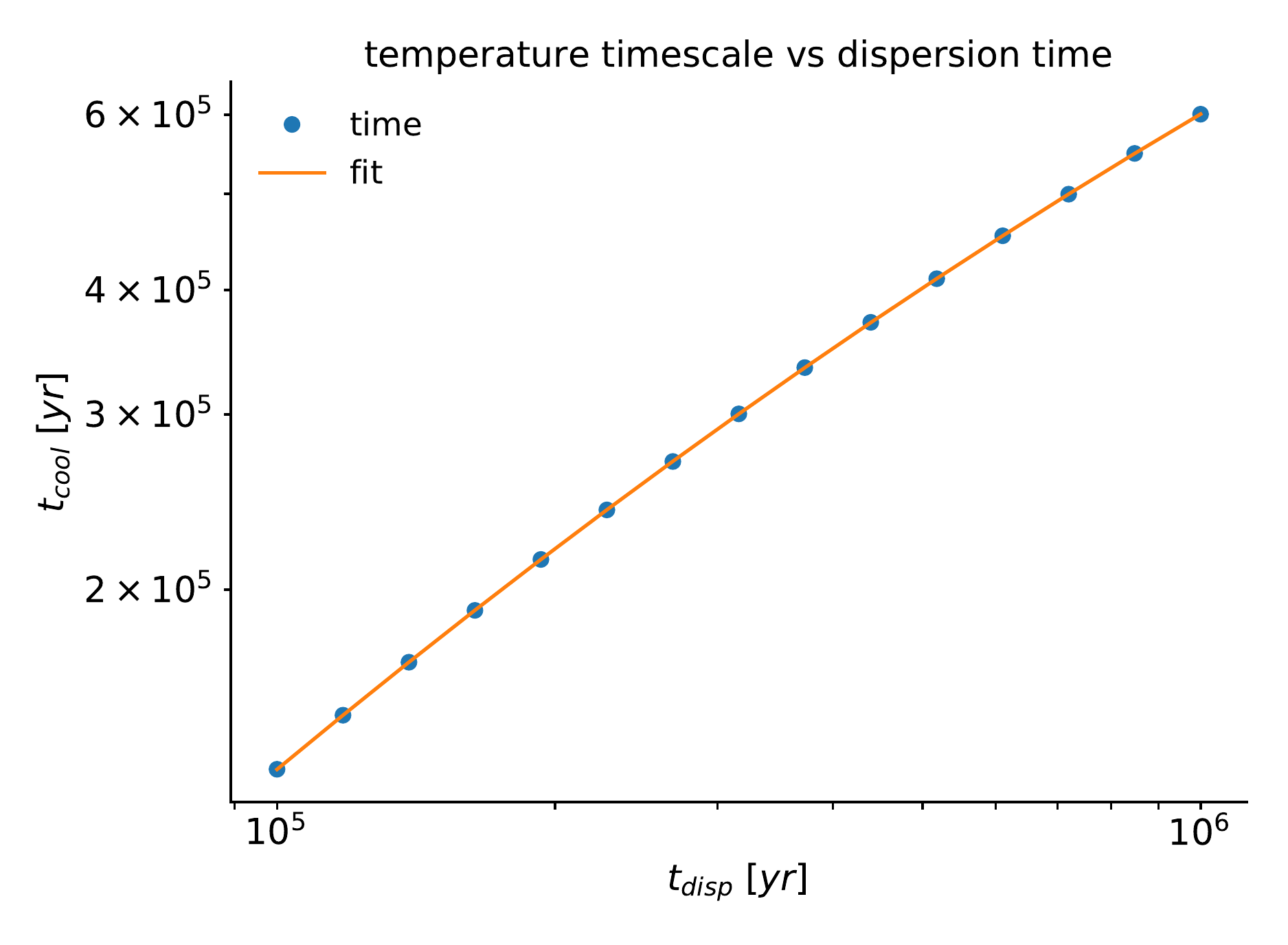}
\caption{Relation between $t_{\rm{cool}}$ and $t_{\rm{disp}}$. The blue dots are the result for 15 different values of $t_{\rm{disp}}$ while the orange line is the fit given by equation \ref{eq:t_temp_fit}.}
\label{fig:plot_t_temp}
\end{figure}

\begin{figure}
\centering
\includegraphics[width=\columnwidth]{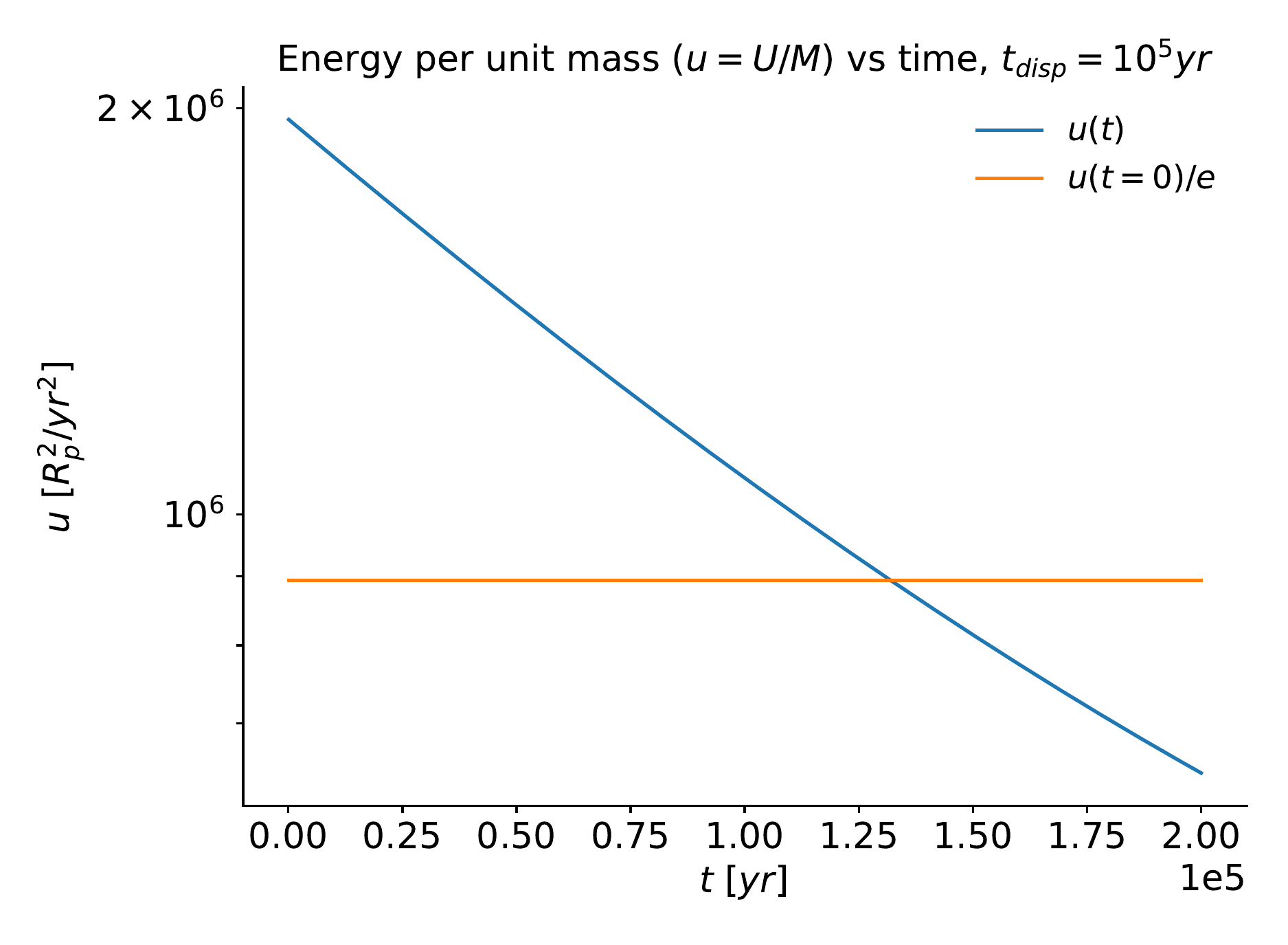}
\caption{Energy per unit mass $u$ in case of $t_{\rm{disp}}=10^5 yr$. The orange line present the initial value for energy divided by $e$ and the blue line is the energy evolution (cooling). The cooling-curve (blue) is nearly exponential.}
\label{fig:t_sample}
\end{figure}

As the optical depth ($\tau$) depends only on $T$ and $\Sigma$, therefore the cooling depends only on how $\Sigma$ varies with time, and it is possible to find a relation between the cooling timescale $t_{\rm{cool}}$ and $t_{\rm{disp}}$ (Fig. \ref{fig:plot_t_temp}). We define $t_{\rm{cool}}$ as the time at which the total internal energy of the disc divided by the total mass of the disc itself ($T\propto U/M$) is $1/e$ of its initial value, as it can be seen in Figure \ref{fig:t_sample}, where it is also clear the exponential nature of the cooling process. This relation is found by fitting the results with $t_{\rm{disp}}$ between $10^5 yr$ and $10^6 yr$:

\begin{equation}\label{eq:t_temp_fit}
log_{10}(t_{\rm{cool}})=-0.11 log_{10}(t_{\rm{disp}})^2+1.9 log_{10}(t_{\rm{disp}})-1.5
\end{equation}
where timescales are in years. We also show this fit in Figure \ref{fig:plot_t_temp}.

The choice of using a ISM mean dust-to-gas ratio is of course an approximation and our results can also be inconsistent when the dust-to-gas ratio is changed. On the other hand, since we obtained a value for $t_{cool}$ that is comparable to $t_{disp}$, the dispersion of the disc would be the dominant effect and therefore, a deeper study would not significantly change our model and outcome.

\subsubsection{Protosatellite formation and evolution}\label{form_and_evo}

\paragraph*{Satellite Formation and Loss}

Once a simulation has started, the code starts to create a new embryo in the position of the dust trap, assuming that the mechanism for dust coagulation is streaming instability \citep{Youdin05}, i.e. a mechanism in which the drag felt by solid particles orbiting in a gas  disc leads to their spontaneous concentration into clumps which can gravitationally collapse. The moonlet formation process starts when these two conditions occur:
\begin{enumerate}
\item The ratio between the solid density and the gas density in the midplane of the dust trap is more than $1$. This condition can occur only if the global dust-to-gas ratio is high enough ($\ge 0.03$ is the threshold in the model, i.e. the initial dust-to-gas ratio should be $\ge 0.005$). This value is given by the profile definition in Section \ref{disc_structure}.
\item The previous proto-moon is far enough, i.e. the dust trap is out of its feeding zone, because of migration.
\end{enumerate}
Once these two conditions have occurred the embryo has to grow to the fixed initial mass ($m_0=10^{-7}M_p$, that is more than two orders of magnitude smaller than individual masses of the Galilean satellites). We use also a formation rate ($\dot{m}_0$) taken from \citet{Drazkowska18}, which we assume to decrease at the same rate as the circumplanetary disc density decreases, i.e. $\dot{m}=\dot{m}_0 e^{-t/\tau}$. Starting from the moment in which the two above-mentioned conditions occur we integrate this formation rate in time until $m=m_0$. At this point the code creates the new protosatellite in the disc. The value for $m_0$ is arbitrary and we tested various $m_0$ to make sure that this initial parameter does not affect results.

The evolution of a protosatellite is stopped in two occasions:
\begin{enumerate}
\item When a protosatellite reaches the inner boundaries of the  disc, then the satellite is considered to be lost into the planet.
\item When two protosatellites intersect their paths the code stops the smallest of the two, even if this very rarely happens. (We are neglecting the possibility that 2 satellites pass each other in 3D.)
\end{enumerate}

Each simulation ends when the total lifetime of the disc is reached, i.e. when $\frac{t}{t_{\rm{disp}}}\sim10$ (see in Section \ref{disc_evolution}).

\paragraph*{Migration}
In our model the orbits of the formed satellites are always considered circular and coplanar \citep{Moraes17} and orbital radii change because of the interaction between the disc and the satellites. In the code we distinguish between type I migration and type II migration. Gap opening separates the two regimes, therefore we use the gap opening parameter $P=\frac{3}{4}\frac{h}{R_H}+\frac{50}{q \,Re} = \frac{3}{4}\frac{c_s}{\Omega_Ka}\left(\frac{M_s}{3M_p}\right)^{-1/3}+50 \alpha\frac{ M_p}{M_s}\left(\frac{c_s}{\Omega_Ka}\right)^2$ from \citet{Crida07}. We consider that type I takes place if $P>1$, otherwise (if $P<1$) type II operates. In the above formula $h$ is the scale-height of the disc, $R_H$ is the Hill radius of the satellite, $q=M_s/M_p$, $Re$ is the Reynolds number, $a$ is the distance from the central planet, $c_s$ is the speed of sound, $\Omega_K$ is the keplerian velocity at the satellite position.

To compute type I migration velocity we use
\begin{equation}
v_r= b_I \frac{M_s \Sigma_g a^3}{M_p^2}\left(\frac{a}{h}\right)^2 \Omega_K
\end{equation}
where $b_I$ is a parameter that is widely used in the migration community and has been computed in different disc conditions \citep{Paardekooper11, Dangelo10, Dittkrist14}. In our code we use the $b_I$ obtained in 3D non-isothermal simulations in \citet{Paardekooper11}, as a function of the disc density, temperature and satellite mass.
One has also to consider the fact that when a satellite is growing, it is also starting to open a partial gap, therefore the gas density is decreasing in the closer Lindblad locations and as a consequence, migration velocity decreases. This is done by multiplying $b_I$ by the value of the gap depth ($0\le$ depth $\le1$) according to the analytic formula of \citet{Duffell15}.

In type II migration, the satellite migrates with the gap, with velocity computed as in \citet{Pringle81}:
\begin{equation}
v_r=-3(\beta_{\Sigma}+\beta_T+2)\frac{\alpha c_s h}{a}
\end{equation}
where $\beta_{\Sigma}=-\frac{dln\Sigma_g}{dlnr}$ and $\beta_T=-\frac{dlnT}{dlnr}$, or $v_r=-\frac{3}{2}\frac{\alpha c_s h}{a}$ in steady state discs, from which it is possible to define a second $b$ parameter, i.e. $b_{II}=-\frac{3}{2}\frac{c_s^4\alpha}{\Omega_K^4a^6\Sigma_g M_s}$. We also want to underline that $b_{II}$ becomes smaller by a factor of $\sim M_{s}/(4 \pi a^2 \Sigma_g)$ when the satellite grows in mass \citep{Syer95} and changes migration regime (from disc-dominated to satellite-dominated). So we modify $b_{II}$ as:

\begin{equation}
b_{II}\;\;\;\to\;\;\;\frac{b_{II}}{1+\frac{M_{s}}{B}}\;\;\;\;\;,\;\;\;\;\;\;B=4\pi a^2 \Sigma_g
\end{equation}

Furthermore, we also considered a smooth transition between type I and type II migration by using a junction function $z$ from \citet{Dittkrist14}:

\begin{equation}
b=z\left(1/P\right)b_I+\left[1-z\left(1/P\right)\right]b_{II}
\end{equation}
where $z(x)=\frac{1}{1+x^{30}}$ and $P$ is the gap opening parameter defined before.

Since the Galilean satellites are found in resonances, we also tried to resonant trapping in our population synthesis.
Actually, the resonance capturing turned out to be a very rare phenomenon in our model because the inner satellite has to slow down significantly for capture to occur, because it is necessary to have converging orbits. The only strong slowing mechanism in our model would be the gap opening, but as we show in Section \ref{sec::mass}, this happens very rarely.

The migration rates are also used to compute the time-steps in the code. In more detail, the time-steps are never longer than $t_{\rm{disp}}/100$ in order not to lose precision on the disc evolution. Moreover, we also impose that a satellite should never move for more than one tenth of a disc cell (i.e. $1R_p/10$) during its migration. As a consequence each timestep is the minimum value between $t_{\rm{disp}}/100$ and $0.1R_p/|v_{mig}|$, computed separately for each migrating satellite.

\paragraph*{Accretion}

While a protosatellite is migrating in the CPD, it also accretes mass from the dust disc. For a very thin dust disc this accretion prescription is \citep{Greenberg91}:

\begin{equation}
\dot{M}_s=2R_s\bar{\Sigma}_s\sqrt{\frac{GM_s}{R_sv_K^2}}v_K=2\left(\frac{R_s}{a} \right)^{1/2} \bar{\Sigma}_s a^2 \left( \frac{M_s}{M_p}\right)^{1/2} \Omega_K
\end{equation}
where $R_s$ is the radius of the satellite. In the formula, we use $\bar{\Sigma}_s$ (that is different from $\Sigma_s$), because it is the average solid density over the entire feeding zone. The radius of the feeding zone is the same order of magnitude as the Hill-radius, i.e. $R_f=2.3 R_H$ \citep{Greenberg91}. 
This value is then multiplied by the gap depth because if the dust is well coupled with the gas (i.e. it is composed by small, $\leq$ mm, grains), then as the satellite grows and opens a gap, there will be less dust around it to accrete.

Once a satellite has accreted the computed mass during a time-step, it is necessary to subtract this mass from the dust disc density. This dust is taken from the feeding zone proportionally to the available mass in each cell: in each point $i$ of the grid within $R_f$ solid density decreases by a value of $\Delta M(i)=\frac{M_{max}(i)}{\sum_{i}^{R_f}M_{max}(i)} dM$ where $M_{max}(i)$ is the mass available in the $i$-th cell. It often happens that a moonlet accretes all the mass available in the feeding zone, reaching its isolation mass.

After a protosatellite has accreted the mass in the feeding zone and created a gap in the dust, the disc tends to use the dust falling from the PPD's vertical influx to reach the equilibrium again (Figure \ref{disc.density}), according to \citet{Drazkowska18}, where the accretion rate onto the central planet, plus the dust lost by satellite accretion in our case, is equal to the dust infall rate. In practice, going back towards the equilibrium configuration means that dust should fill the dust gaps left by satellite accretion. In the population synthesis, we model this refilling mechanism assuming a typical timescale $t_{\rm{refilling}}$ for this process, that could cover a wide range of values, since the dust-to-gas ratio within the vertical influx of material is unknown. In our model the CPD gains mass in the following way from the vertical influx:
\begin{equation}
\Delta \Sigma_s=\begin{cases}
\frac{\hat{\Sigma}_s-\Sigma_s}{t_{\rm{refilling}}}dt &dt\le t_{\rm{refilling}}\\
\hat{\Sigma}_s-\Sigma_s &dt>t_{\rm{refilling}}
\end{cases}
\end{equation}
where $dt$ stands for the time-step, ${\Sigma}_s$ is the current solid density and $\hat{\Sigma}_s$ is the value that the solid density would have if there was not accretion and consequent depletion.

The timescale of this process is not well constrained, because it strongly depends, for instance, on the amount of dust that fall into the CPD from the PPD, that can be either very fast, with $t_{\rm{refilling}} \sim 10^2 yr$, or very slow, with $t_{\rm{refilling}} \sim 10^6 yr$.

\subsubsection{Population synthesis}\label{pop_synthesis}

The last module of the code allows to run the semi-analytical algorithm with a population synthesis approach. 
The idea of population synthesis is to explore a range of the unconstrained parameters, trying all the different combinations between them and in the end to compare the results, individually or grouped. The parameters we vary in the population synthesis are:

\begin{itemize}

\item the dust-to-gas ratio in $(0.03,0.50)$, changing only the dust component
\item the CPD dispersion timescale: $t_{\rm{disp}}$ in $(10^5,10^6) yr$ 
\item the dust refilling timescale: $t_{\rm{refilling}}$ in $(10^2,10^6) yr$
\end{itemize}

In random cases we distribute $t_{\rm{disp}}$ exponentially, as described by \citet{Fedele10}, while we distribute dust-to-gas ratio and $t_{\rm{refilling}}$ logarithmically. Furthermore, we vary when the simulation begins, in order to have different initial conditions in temperature and density profiles of the disc. The simulation can start anytime between $0$ and $t_{\rm{disp}}/2$.

In principle one can set lower dust-to-gas ratios but since streaming instability is only occurring when the dust-to-gas ratio is $>0.03$ we did not consider those low dust-to-gas ratio cases in our results. There will be, of courses, cases with dust-to-gas ratios $<0.03$ but estimating their number would be possible only when the global dust-to-gas ratio distribution will be clear. For instance, calling the dust-to-gas ratio variable $x$, if we assume a logarithmic probability distribution within $0.01<x<0.50$, i.e. $dp/dx\propto 1/x$, and we extend the distribution in order to go to $0$ for low dust-to-gas ratios (for example $dp/dx \propto 100x$ in $0<x<0.01$ seems reasonable), we find that about $35\%$ of the cases have dust-to-gas ratio $<0.03$.

\begin{figure}
\includegraphics[width=1.1\columnwidth]{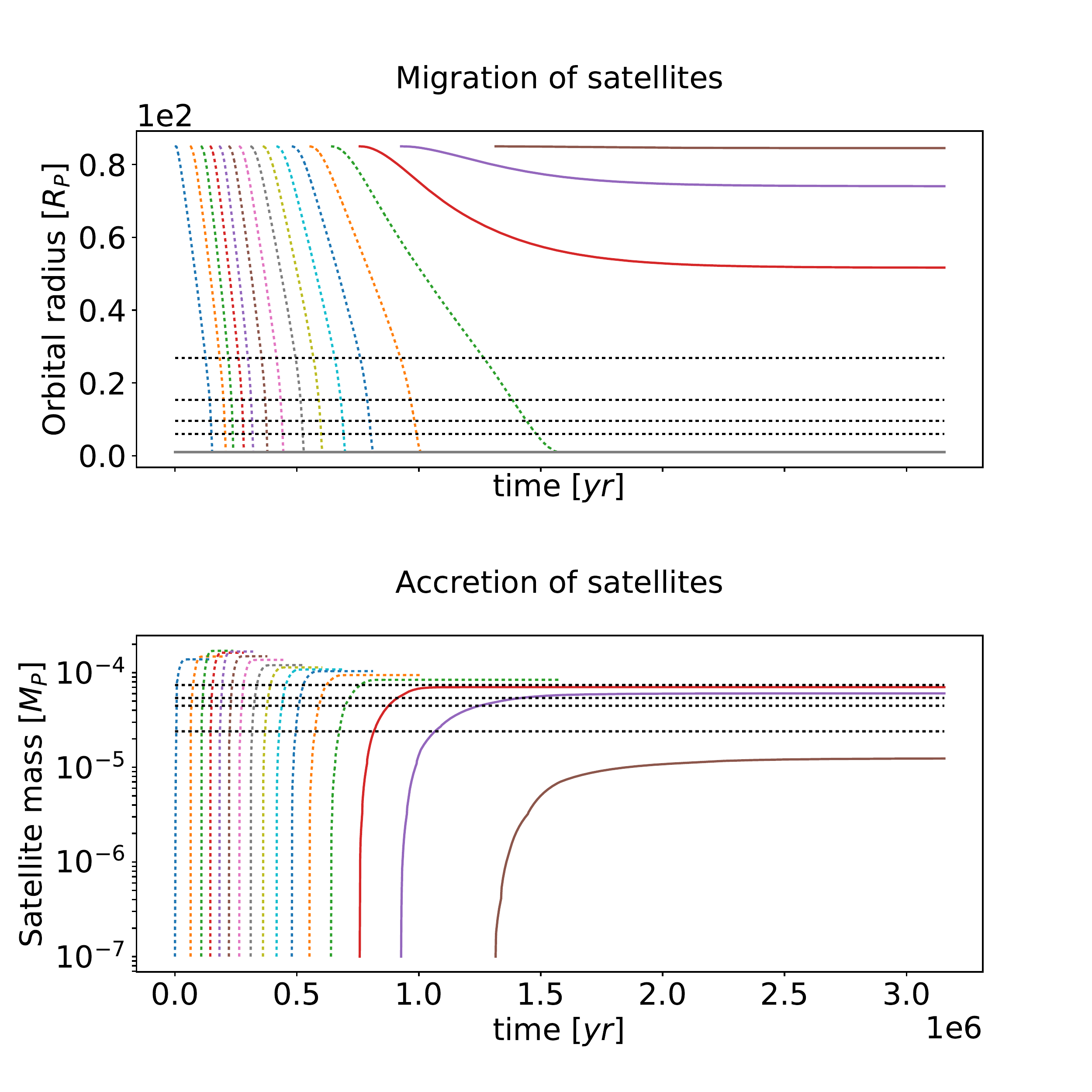}
\caption{Evolution of satellites in a system with dust-to-gas ratio $=0.1$, $t_{\rm{disp}}=10^5$ yr and $t_{\rm{refilling}}=2\times 10^4$ yr. Solid lines are the surviving satellites, dashed lines are lost ones.}
\label{fig:plot}
\end{figure}

One could also vary other parameters, such as the initial embryos mass or the type I migration formula used. We tested these, but this did not change the results much, therefore we kept them fixed as described in the previous sections. We show in Figure \ref{fig:plot} how the results of a single run look, with satellites growing, being lost and migrating within a CPD. We also note that there are parameters we kept fixed to be consistent with the hydrodynamic simulation, but they could have been varied too. 

\section{Results}\label{results}

In our work we used two kinds of approaches for population synthesis: the first one consists in running twenty-thousands of different simulations with randomizing the three initial parameters described in the previous section. The second approach is controlling a value for a single parameter, and let the other two vary randomly. The first approach allows to have a general understanding of the outcomes, respecting parameter distribution (especially the exponential distribution of $t_{\rm{disp}}$, that is an observational constraint), while the second approach allows to understand how a single parameter affects the results.

\subsection{Survival timescale of the last generation of satellites}\label{t_LG}

\begin{figure}
\centering
\includegraphics[width=\columnwidth]{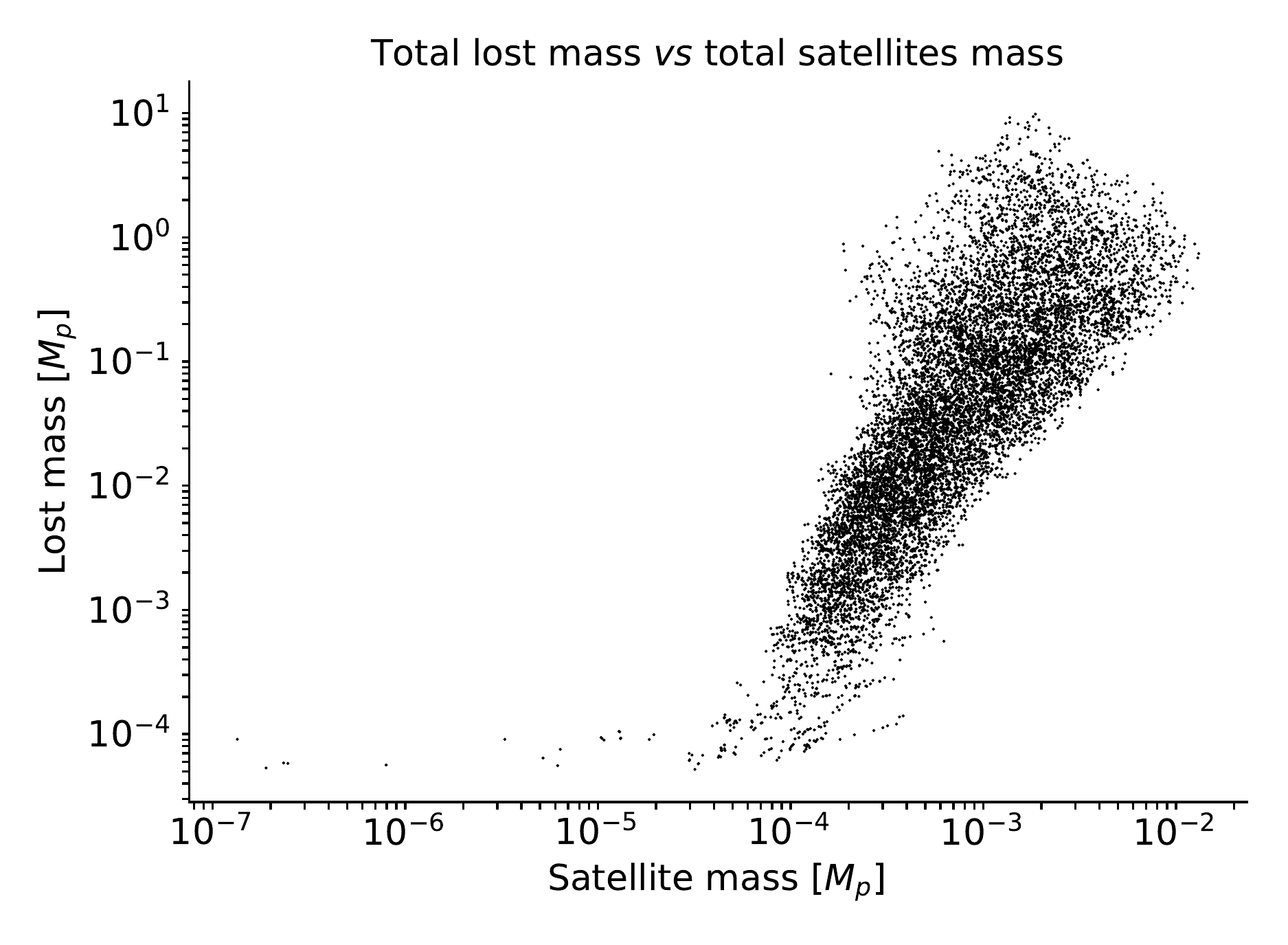}
\caption{The total lost satellite mass that is polluting the planet versus the total mass of the surviving satellites (see Section \ref{sec::mass}). For $M_p=M_J$, the median value of lost satellites is 15 $ M_{\oplus}$, but the scatter is large.}
\label{fig:lost_mass_vs_survived_mass}
\end{figure}

Due to the fact that the moonlets migrate inwards in the disc (see Section \ref{sec::mass}), and it is often assumed that there is no cavity between the planet and the CPD (see e.g. \citealt{OM16} and \citealt{SzM17}), many (even a dozen of) satellites are lost into the planet during disc evolution and therefore only the latest set of moons will survive when the CPD (and PPD) dissipates. This is called sequential satellite formation, that was already suggested in e.g. \citet{Canup02}. These lost satellites pollute the envelope of the forming giant planet, increasing the metallicity of the gas-giant. Given that Jupiter's atmosphere is enriched $\sim 2$ times compared to the protosolar values (e.g. \citealt{Bolton17}), these lost satellites (and the continuous dust drift/migration) might contribute to this overabundance of heavy elements.
Therefore, we computed the mass, what the lost satellites bring into Jupiter: we found a distribution with a median value of $\simeq 5\times 10^{-2} M_J\simeq 15 M_{\oplus}$ (Figure \ref{fig:lost_mass_vs_survived_mass}), but the scatter is large around this value.

\begin{figure}
\includegraphics[width=\columnwidth]{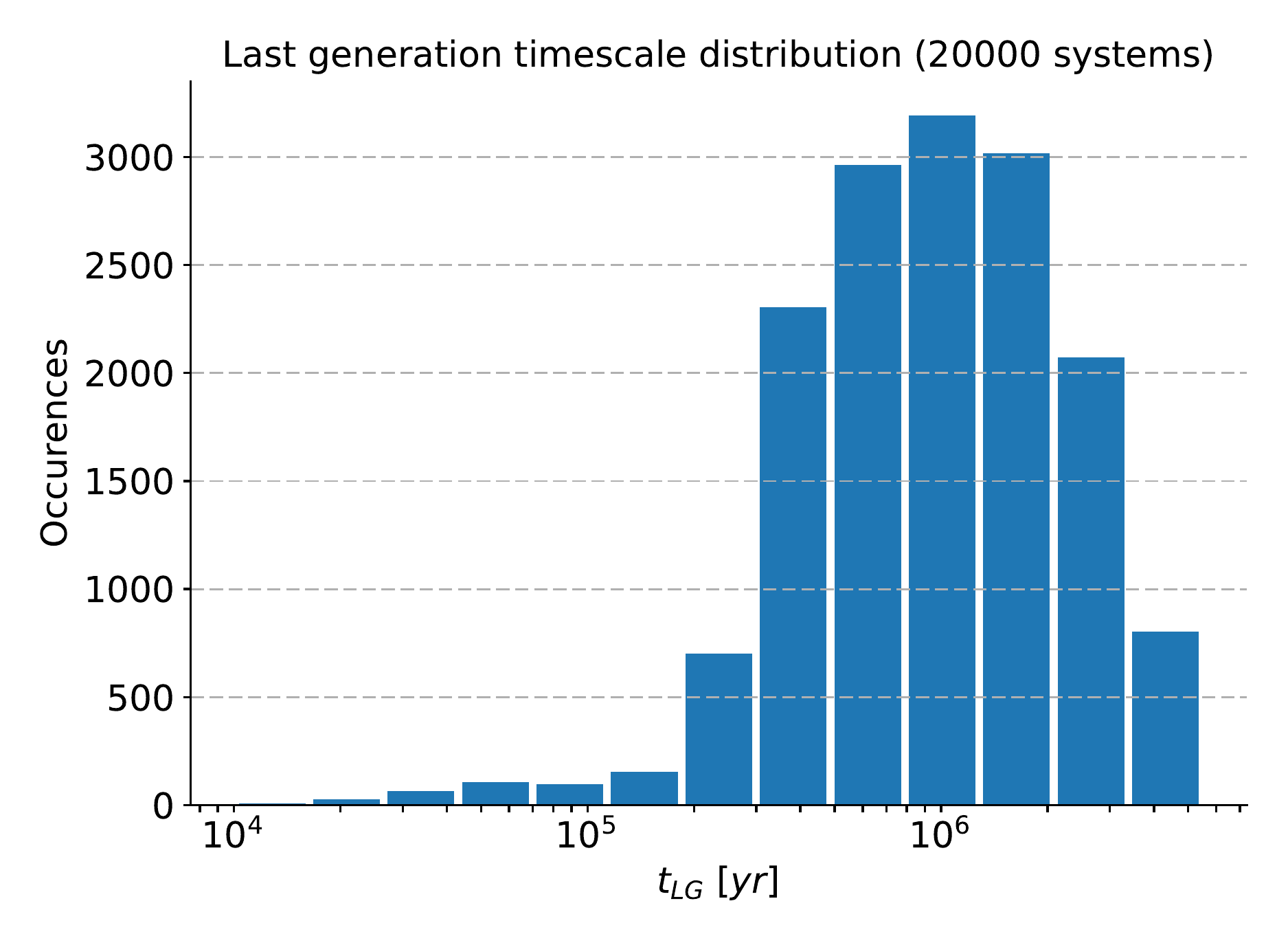}
\caption{Histogram of all the last generation survival timescales for systems in which at least one satellite survives.}
\label{fig:results:t_LG}
\end{figure}

Proceeding with the first type of population synthesis approach (see the first paragraph of Section \ref{results}) it is possible to study the general behavior of forming satellite-systems. Running 20000 simulations, we found $4467$ ($22.34\%$) systems in which there are not survived satellites, $325$ ($1.62\%$) systems in which all satellites survive and $15208$ ($76.04\%$) systems in which at least one moon is lost but at the same time at least one satellite survives.
This fact is confirmed in Figure \ref{fig:results:t_LG}, where we show the distribution of what we call \textit{last generation timescale} (hereafter $t_{LG}$) for $20000$ satellites. $t_{LG}$ is defined as the amount of time that a system takes, starting from the beginning of the simulation, to form the last generation of surviving satellites. The figure indicates that most of the surviving satellites form between $2\times 10^5$ and $5\times 10^6$ years ($93\%$ of the cases).

\begin{figure}
\includegraphics[width=\columnwidth]{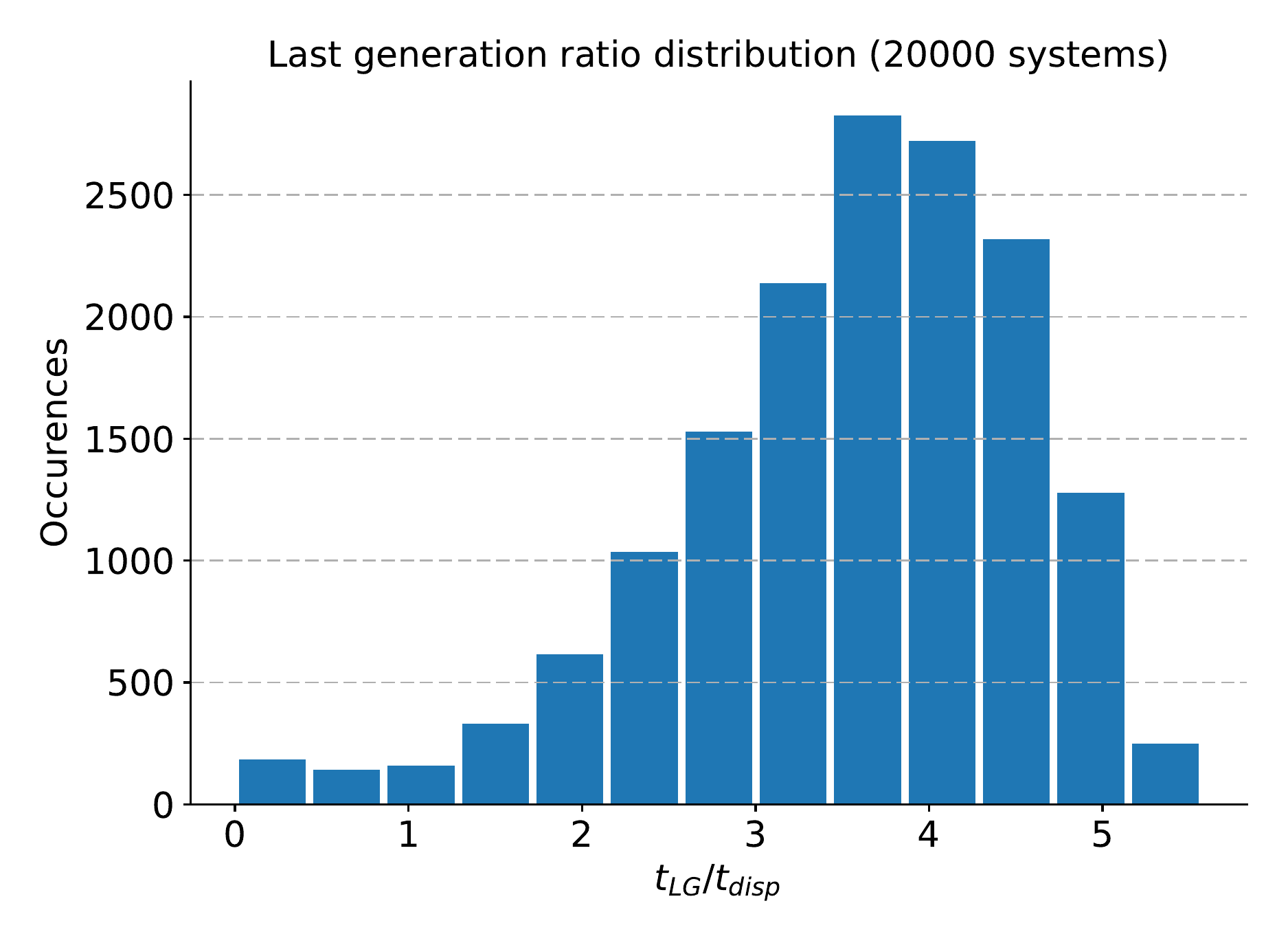}
\caption{Histogram of all the last generation survival timescales ($t_{LG}$) for systems in which at least one satellite survives, divided by $t_{\rm{disp}}$ (ratio).}
\label{fig:results:LG_ratio}
\end{figure}

In fact, using years, as in Figure \ref{fig:results:t_LG}, could be misleading, since there is an order of magnitude of difference between the fastest evolving discs ($t_{\rm{disp}}=10^5 yr$) and the slowest one ($t_{\rm{disp}}=10^6 yr$). Calculating the distribution of $t_{LG}/t_{\rm{disp}}$ allows to better study how late surviving satellites form compared to the dispersion timescale of the disc and as a consequence, to the total lifetime ($t_{life}\simeq 10 t_{\rm{disp}}$). It is clear that they form very late in the system evolution, even after $5$ dispersion timescales, i.e. after $50\%$ of the total lifetime of the  disc (see Fig. \ref{fig:results:LG_ratio}), when usually discs are already very poor of gas and dust, having about $0.5\%$ of the initial mass. Here we always refer to the dispersion timescale because it is the fundamental quantity that defines the evolution of a disc (e.g. $\Sigma(t) \propto exp(-t/t_{\rm{disp}})$). 

\begin{figure}
\centering
\includegraphics[width=\columnwidth]{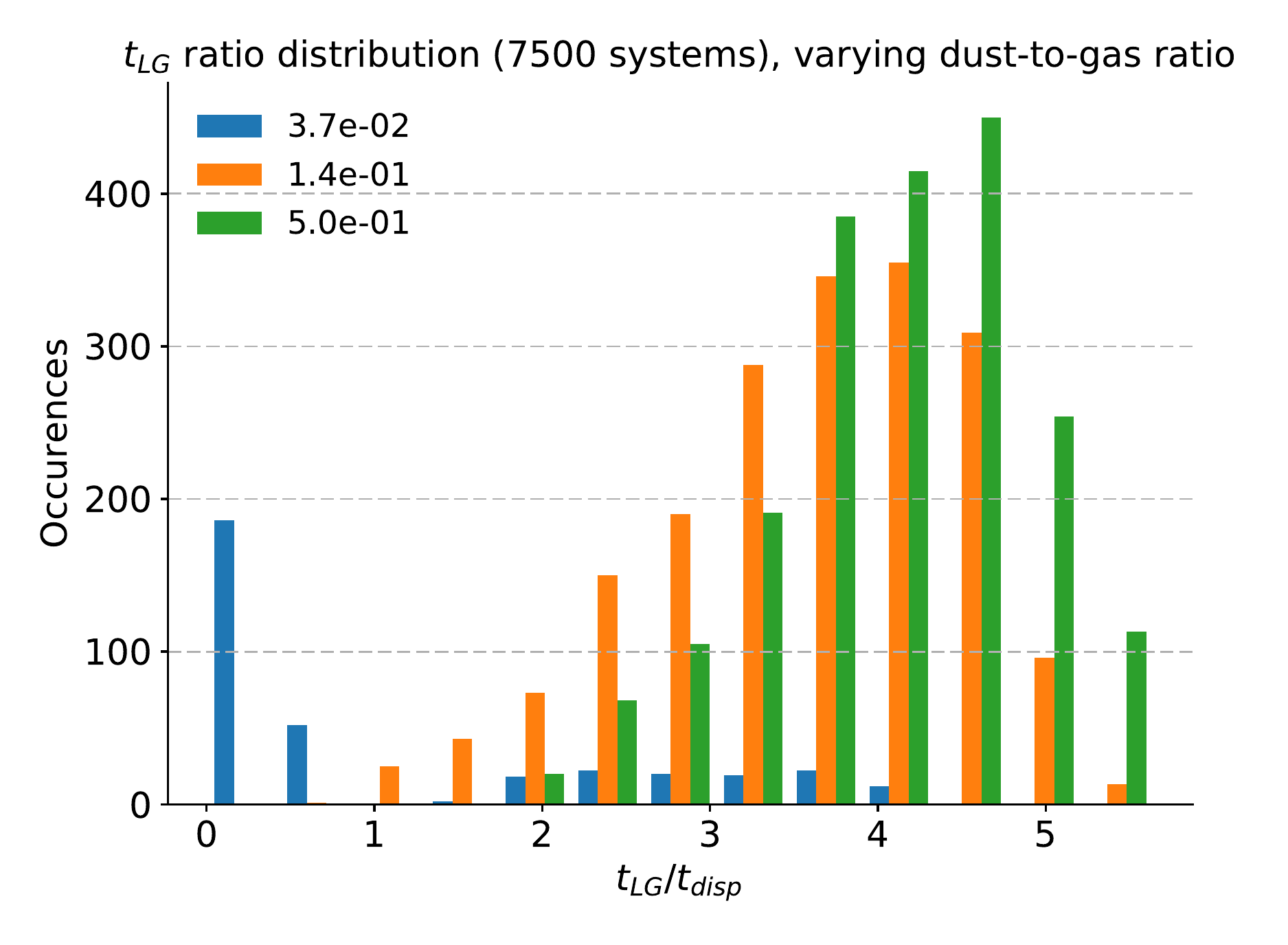}
\includegraphics[width=\columnwidth]{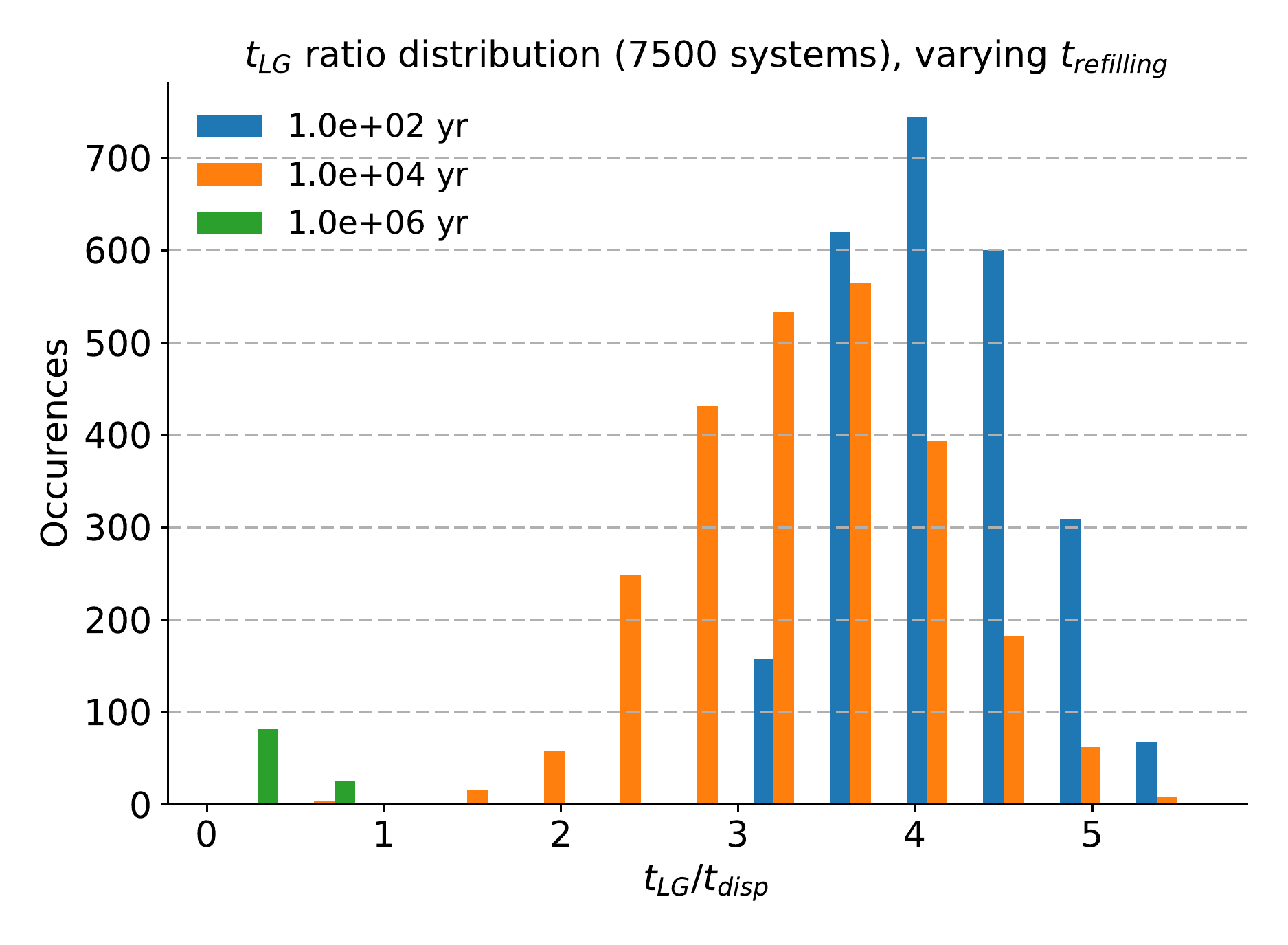}
\includegraphics[width=\columnwidth]{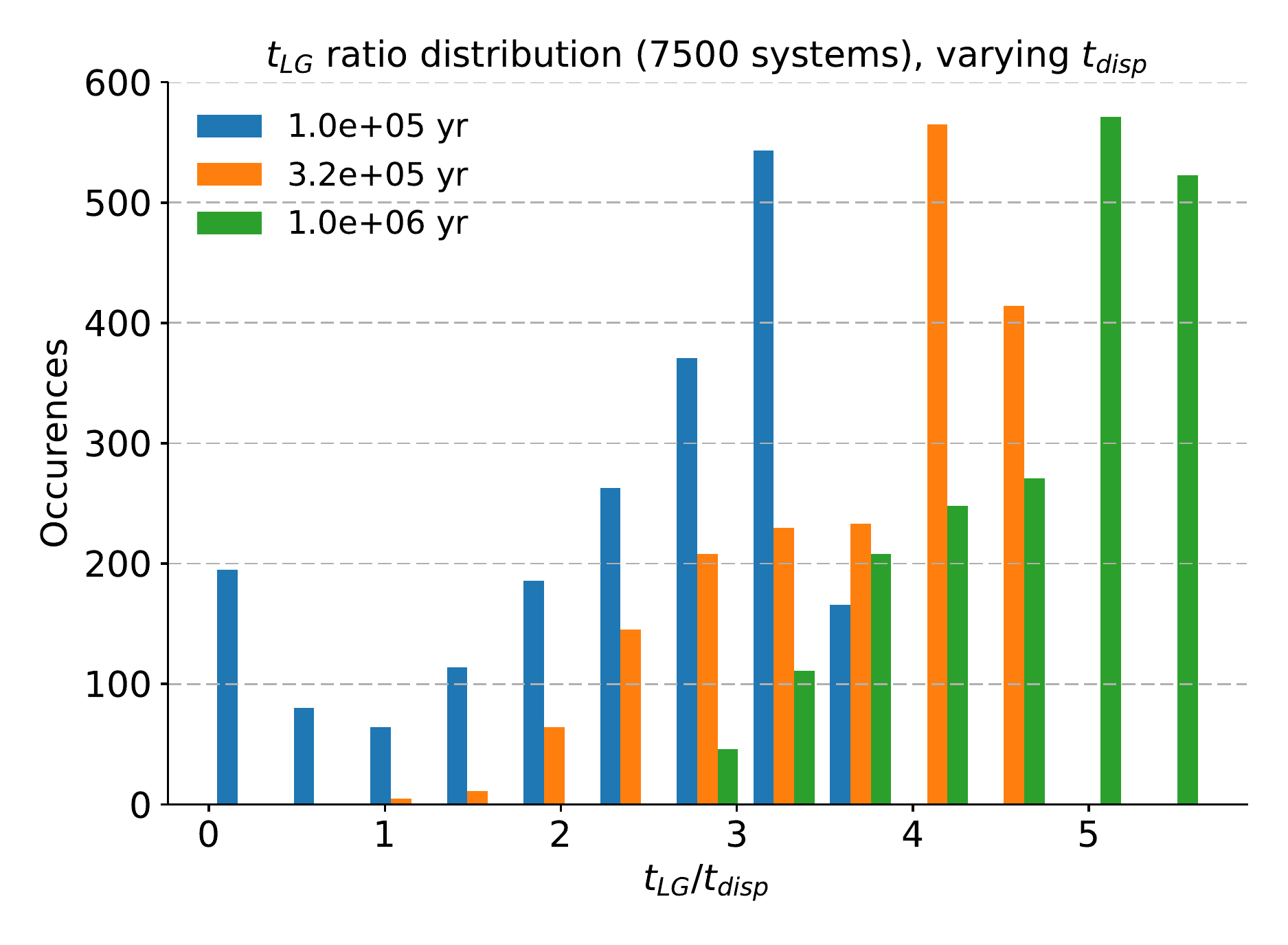}
\caption{Distributions of $t_{LG}/t_{\rm{disp}}$  varying dust-to-gas ratio (top panel), varying the refilling timescale (middle panel), and changing the disc dispersion timescale (lower panel). 2500 simulations were run for each value (7500 simulations in total).}
\label{fig:LG_ratio_varying}
\end{figure}

It is also possible to analyse the model with the second type of population synthesis (see the first paragraph of Section \ref{results}), in order to understand how different parameters affect results. Considering $t_{LG}/t_{\rm{disp}}$ as the most significant parameter to investigate, we performed the study presented in Figure \ref{fig:LG_ratio_varying}. The first thing that these plots show is that $t_{LG}/t_{\rm{disp}}$, and as a consequence the sequential formation mechanism, is highly dependent on all the parameters we chose in our population synthesis model. For example in the upper panel it is shown that satellites that survives have to form earlier, when less dust is available. This is because when the disc is poor of dust it would be difficult to reach streaming instability conditions in the dust trap, then whether the first generation of satellites survives (then we have very short $t_{LG}$s) or the first generation is lost into the central planet, following generations can not survive.

The second dependence (i.e. $t_{\rm{refilling}}$, second panel on Figure \ref{fig:LG_ratio_varying}), shows that the faster the refilling mechanism is, the later the moons form. This is again related to streaming instability conditions in the dust trap because if refilling is efficient the  disc would be able to provide enough dust to the dust trap to form a lot of satellites even at later stages.

In the lower panel of Figure \ref{fig:LG_ratio_varying} the dependence on $t_{\rm{disp}}$ is plotted. According to this, satellites form later if $t_{\rm{disp}}$ is longer. 
This is not related to streaming instability conditions as before, but simply if the dissipation time is shorter, the gas density decreases quicker and the survived generation forms earlier. because migration and accretion stop earlier. This means that the timescales of the model are not linear with $t_{disp}$. This is clear because migration and accretion timescales of a single satellite just slightly depend on the dissipation timescale.

\subsection{Formation timescales}\label{t_form}

Since we still do not really know on what timescales the Galilean satellites were formed, i.e. how much time a satellite takes to form starting from the formation of its initial embryo, the population synthesis can give a hint about this. Formation timescales have an impact on the structure and composition of the moons, or in reverse, the internal structures of Galilean satellites provide some constraints on the formation timescale. The three inner satellites show a differentiated structure \citep{Anderson96,Anderson98,Anderson01}, while Callisto, on the other hand, is not completely differentiated \citep{Sohl02}. Differentiation occurs when a satellites (or a body, in general) melts because of the energy received from gas interactions, satellitesimals collisions, etc. When this happens heavy elements are allowed to sink toward the centre of the satellite, creating different layers. The structure of Callisto gives a first caveat about its evolution, i.e. some believe that its formation timescale could not be shorter than $\sim10^5 yr$ because otherwise collisions and accretion would have transferred energy at a rate high enough to have complete melting \citep{Canup02, Stevenson86}. However we have very little knowledge on how the heating/cooling processes worked in the circumplanetary disc that created this moon, nor, where inside the disc Callisto has formed and how its migrated.

\begin{figure}
\centering
\includegraphics[width=\columnwidth]{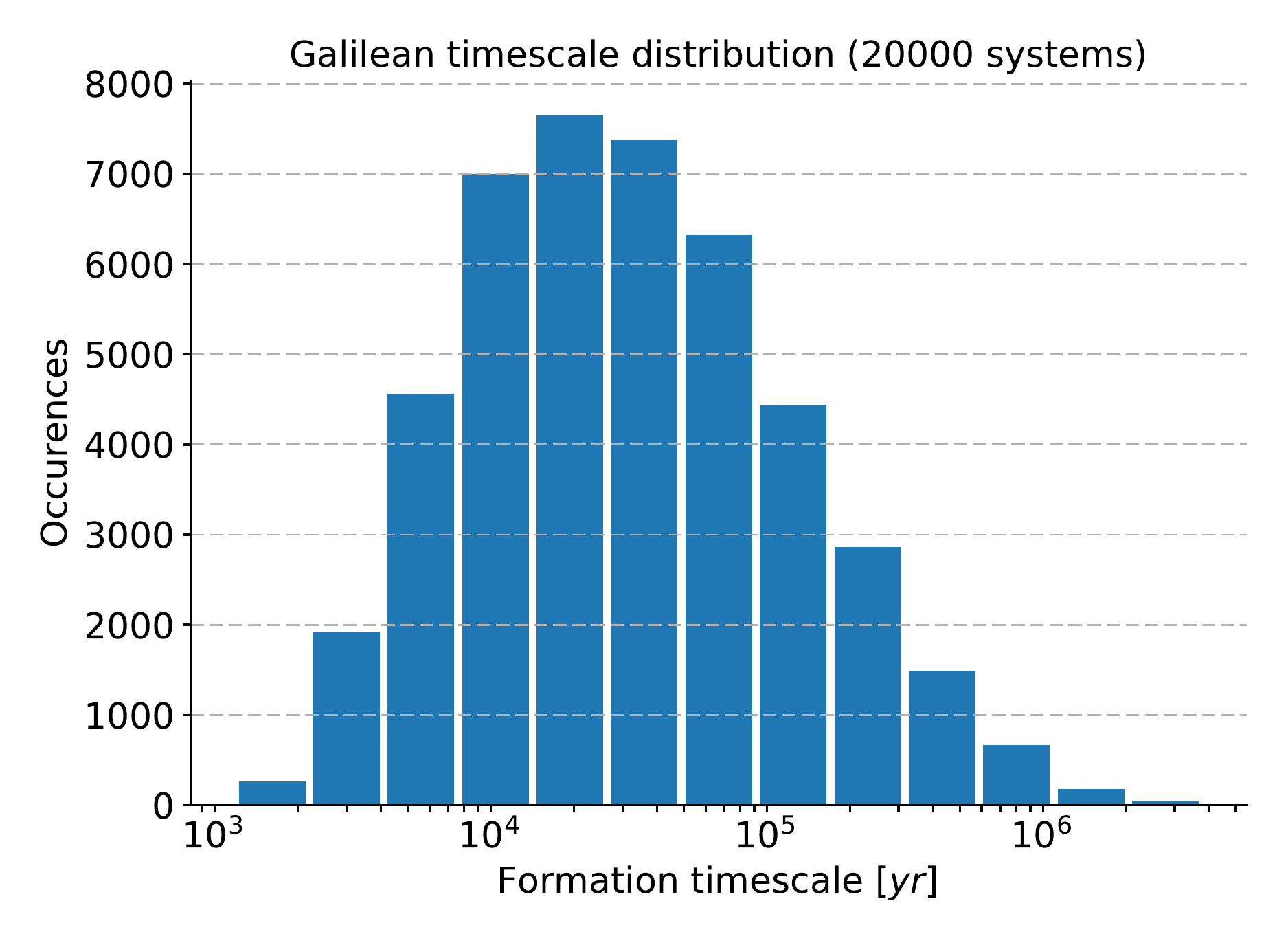}
\caption{Histogram of all the formation timescales, that distribute with a peak around $2\times10^4$ yr, with cases in which satellites form even faster than $2-3 \times 10^3$ yr.}
\label{fig:formation_histogram}
\end{figure}

In all the simulations it is possible to look at the time needed by any survived satellite to grow to a typical Galilean mass (we chose Europa's mass as a benchmark) and see how these timescales, that we call \textit{formation timescale}, distribute, leaving out the (few) satellites that do not grow up to Europa's mass. 
The formation timescale distribution is shown in Figure \ref{fig:formation_histogram} in which the distribution has a maximum between $10^4$ and $10^5$ yr with cases down to $10^3$ yr (about $20\%$ of the population forms less than $10^4 yr$). This means that satellites can also form very quickly, compared to terrestrial planet formation timescales. This is especially true if the dust-to-gas ratio is high enough in the CPD, the refilling mechanism is efficient and  disc dispersion is fast. Previous models, as in \citet{Canup02}, predicted quite long timescales, because they did not consider a strong influx from the PPD and, as a consequence, a dust refilling mechanism, instead they just have a low influx rate from the PPD ($<10^{-6} M_J/yr$) in order to have low temperature and long accretion rates for satellites, to prevent melting and differentiation.

\begin{figure}
\centering
\includegraphics[width=\columnwidth]{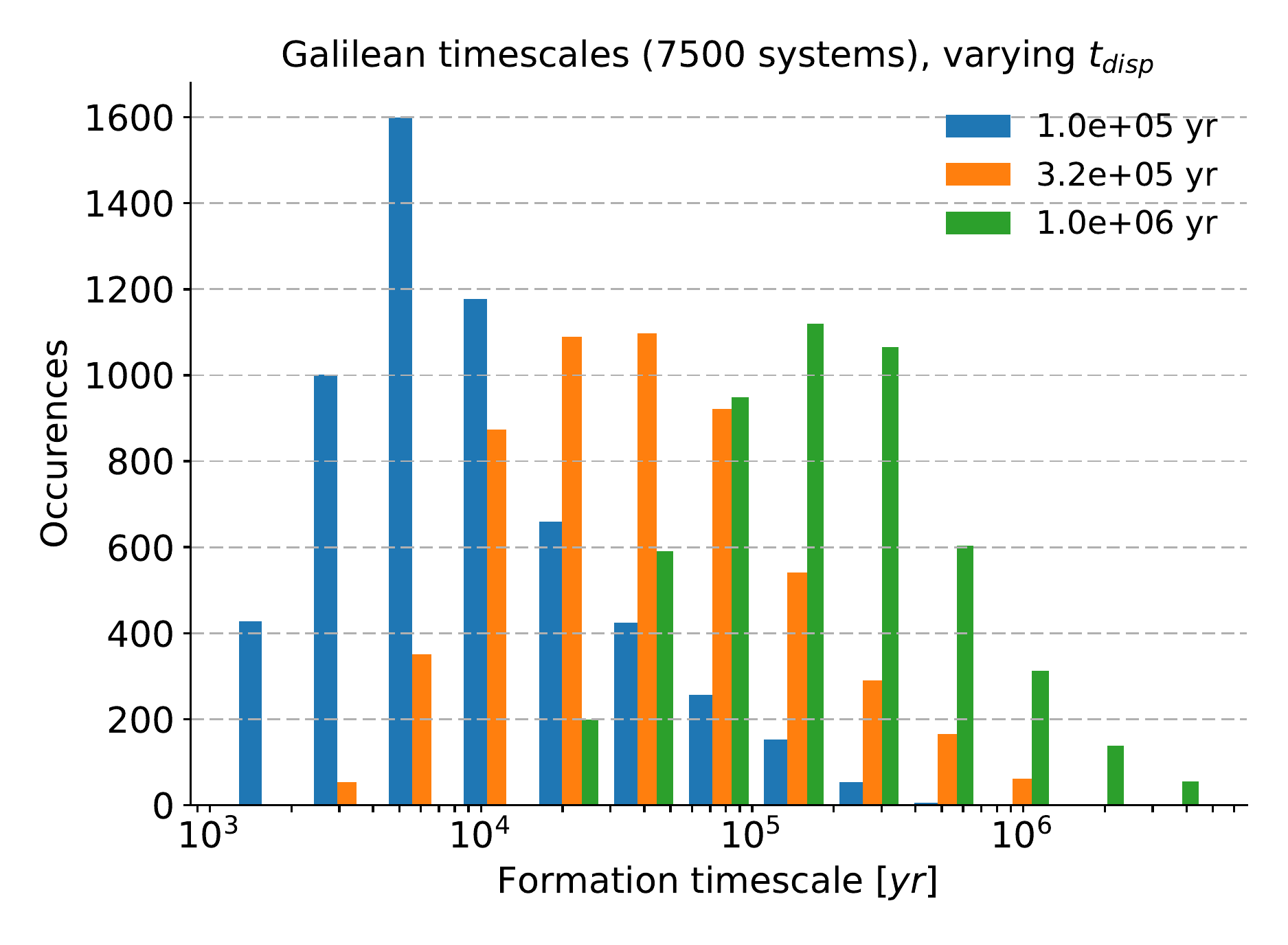}
\caption{Distribution of the formation timescale while varying the value of $t_{\rm{disp}}$, running 2500 simulations for each value (7500 simulations in total).}
\label{fig:t_form_varying}
\end{figure}

Our results on the formation timescale do not disagree with the $\sim10^5 yr$ timescale cited before, because a part of the population is forming on such a long timescale. 
We have also checked the dependence of the formation timescale on $t_{\rm{refilling}}$, on $t_{\rm{disp}}$, and on the dust-to-gas ratio. Satellites of course form faster and bigger when there is more available dust and/or when refilling is efficient. However, a non-trivial dependence is that on $t_{\rm{disp}}$ because it is not possible to link it simply to a general availability of dust or to the efficiency of accretion. The dependence is more related to $t_{LG}/t_{\rm{disp}}$, exactly as we described in Section \ref{t_LG}. According to this, if $t_{\rm{disp}}$ is longer, then $t_{LG}/t_{\rm{disp}}$ is longer and the formation process is slower because there is less dust available. This is because the dust density depends exponentially on $t/t_{\rm{disp}}$, see Section \ref{disc_evolution}, and the same mechanism applies the other way round. In Figure \ref{fig:t_form_varying}, the dependence of the formation timescale on the disc dispersion timescale is summarized.

\subsection{The mass distribution of the satellites}\label{sec::mass}

\begin{figure}
\centering
\includegraphics[width=\columnwidth]{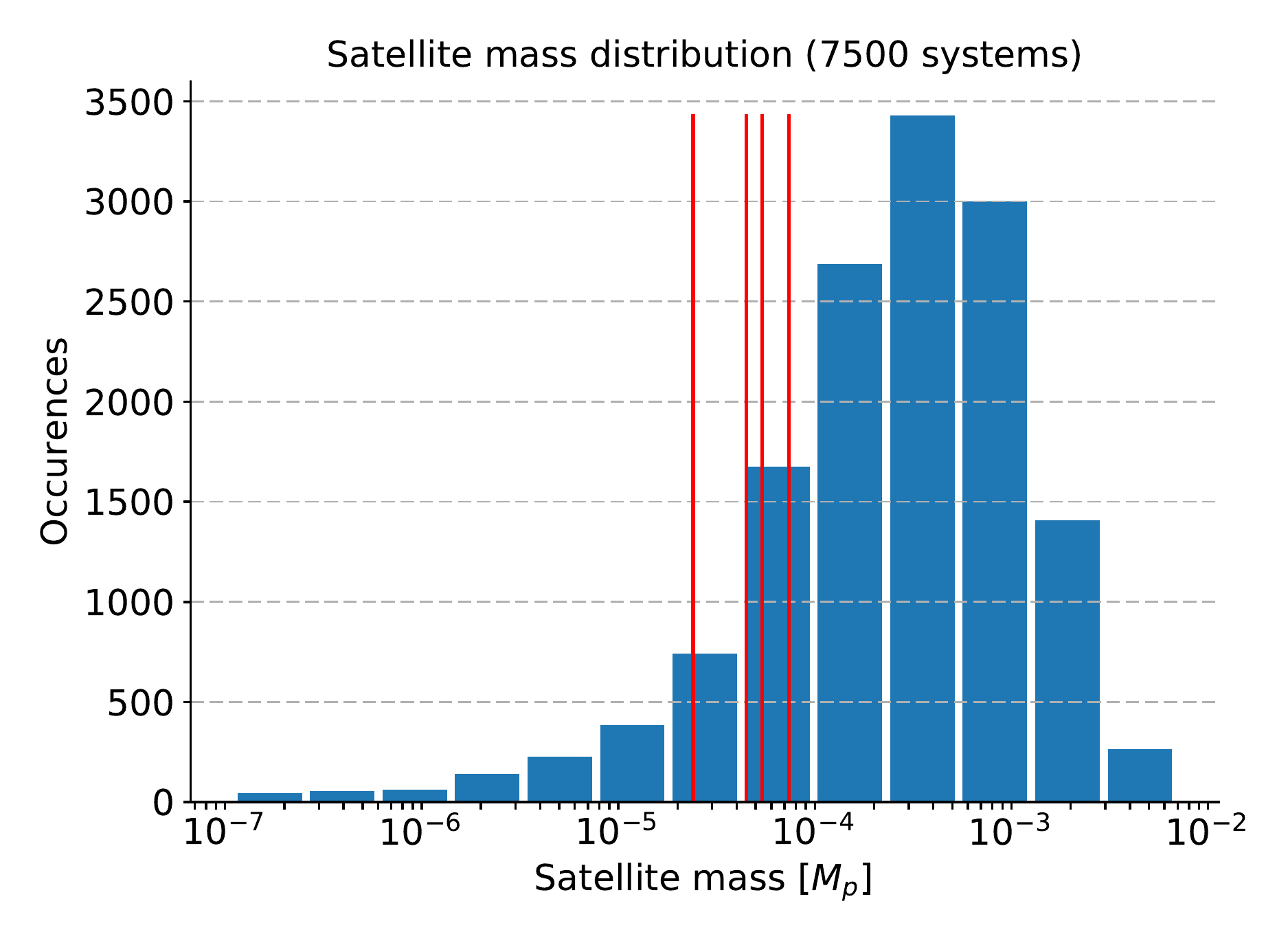}
\caption{Satellite-mass distribution. The peak can be found between $10^{-4}$ and $10^{-3}$ $M_p$, i.e. larger than the Galilean masses, in fact almost at Earth-mass. Red lines indicates the masses of the four Galilean satellites individually.}
\label{fig:histogram_masses}
\end{figure}

The mass distribution of surviving satellites is shown in Figure \ref{fig:histogram_masses}, with red vertical lines representing the masses of the four Galilean moons. According to this histogram, the population spreads between $10^{-7} M_p$ (i.e. the initial mass of embryos), and $10^{-2} M_p$. The peak of the distribution is between $10^{-4}$ and $10^{-3}$ $M_p$, which is higher than Galilean masses, often reaching Earth-mass. Only $\sim 10 \%$ of the population has a mass similar to Galilean ones. 

\begin{figure}
\centering
\includegraphics[width=\columnwidth]{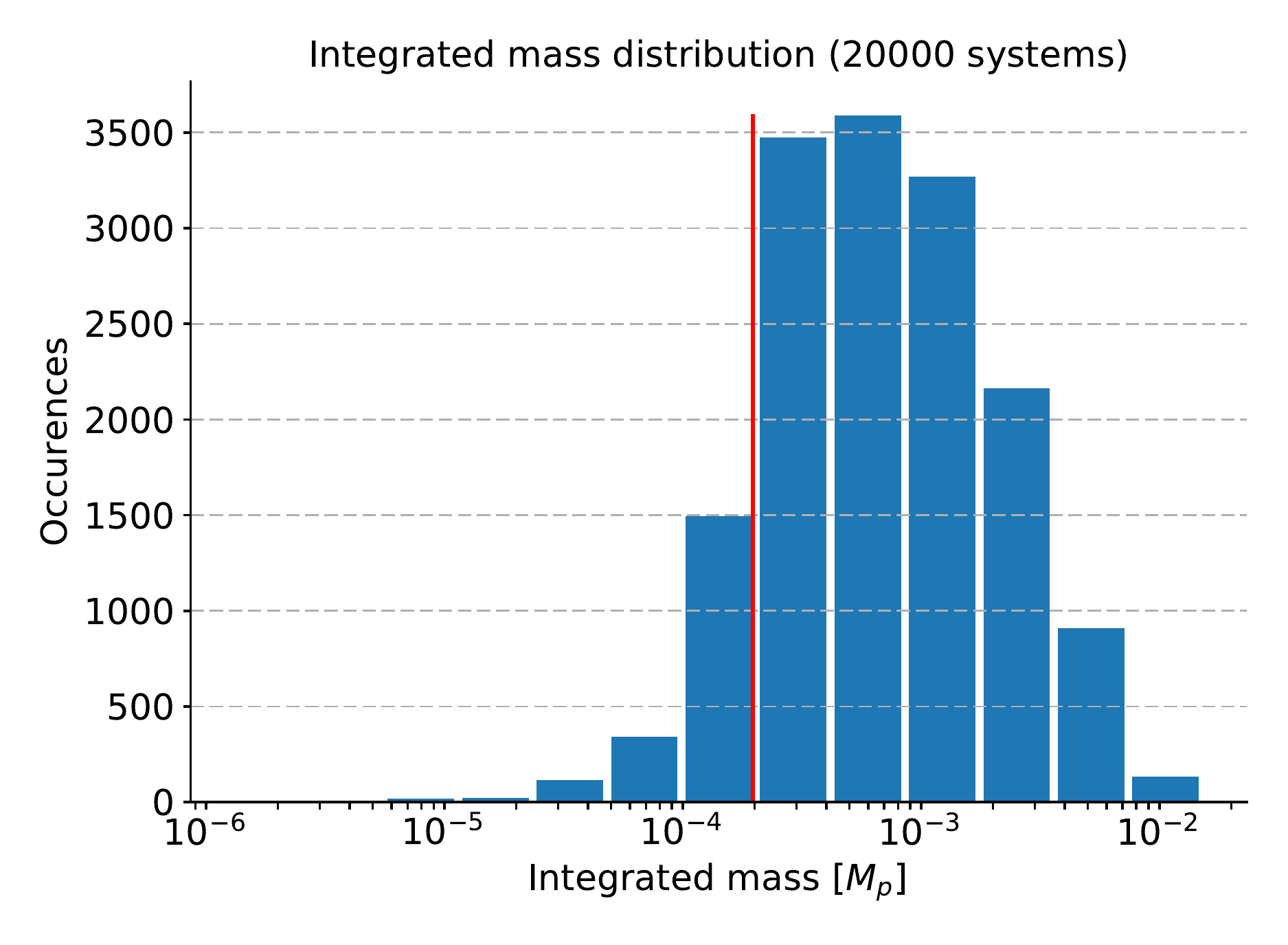}
\caption{Satellites integrated mass distribution. It has a peak between $10^{-4}$ and $10^{-3}$ $M_p$, while the upper limit is about $10^{-1} M_p$. The distribution is symmetric. Red line is the Galilean satellites'  integrated mass ($\simeq 2\times 10^{-4}M_p$).}
\label{fig:histogram_integrated_masses}
\end{figure}

It was pointed out e.g. in \citet{Canup02} that the ratios between the integrated masses ($M_{int}$) of the moons of Jupiter and Saturn, and the mass of the relative planets themselves, are the same: $M_{int}/M_p=2\times 10^{-4} $. The authors there discuss the possibility, whether this is coming from physics somehow, whether the CPD-mass is only based on the planetary mass. Recent hydrodynamic simulations have shown, however, that not only the planetary mass sets the CPD-mass, but also the PPD-mass, since the latter continuously feeds the former, hence the more massive PPD will produce a more massive CPD around the same massive planet \citep{Szulagyi17}. To check those results with population synthesis, in Figure \ref{fig:histogram_integrated_masses} we plotted the histogram of the integrated mass of moons in each individual system of the population. The vertical red line again highlights the Galilean integrated satellite mass: ($2\times 10^{-4} M_{\rm{Jupiter}}$). From the Figure it can be concluded that the integrated mass of satellites has a wide distribution, there is no hint for any physical law producing a peak at $M_{int}=2\times 10^{-4} M_p$, or at any other particular mass. We therefore conclude, that it is just a coincidence, why the integrated mass of satellites of Jupiter and Saturn are $2\times 10^{-4} M_{p}$.

We also checked in how many cases, out of the total 20 thousands, we get systems with 3 or 4 satellites with a total mass between $10^{-4} M_p$ and $4\times 10^{-4} M_p$, i.e. systems that have masses similar to the Galilean ones. We found that about 4200 systems have such characteristic, i.e. about $21\%$ of the cases. It is easier to have such systems when the dispersion time of the disc is as long as possible ($\to 10^6 yr$) and the refilling timescale is between $10^4$ and $10^5$ year, while in those cases the value of dust-to-gas ratio can vary in a very wide range (from $5\%$ to $20\%$).

\begin{figure}
\centering
\includegraphics[width=\columnwidth]{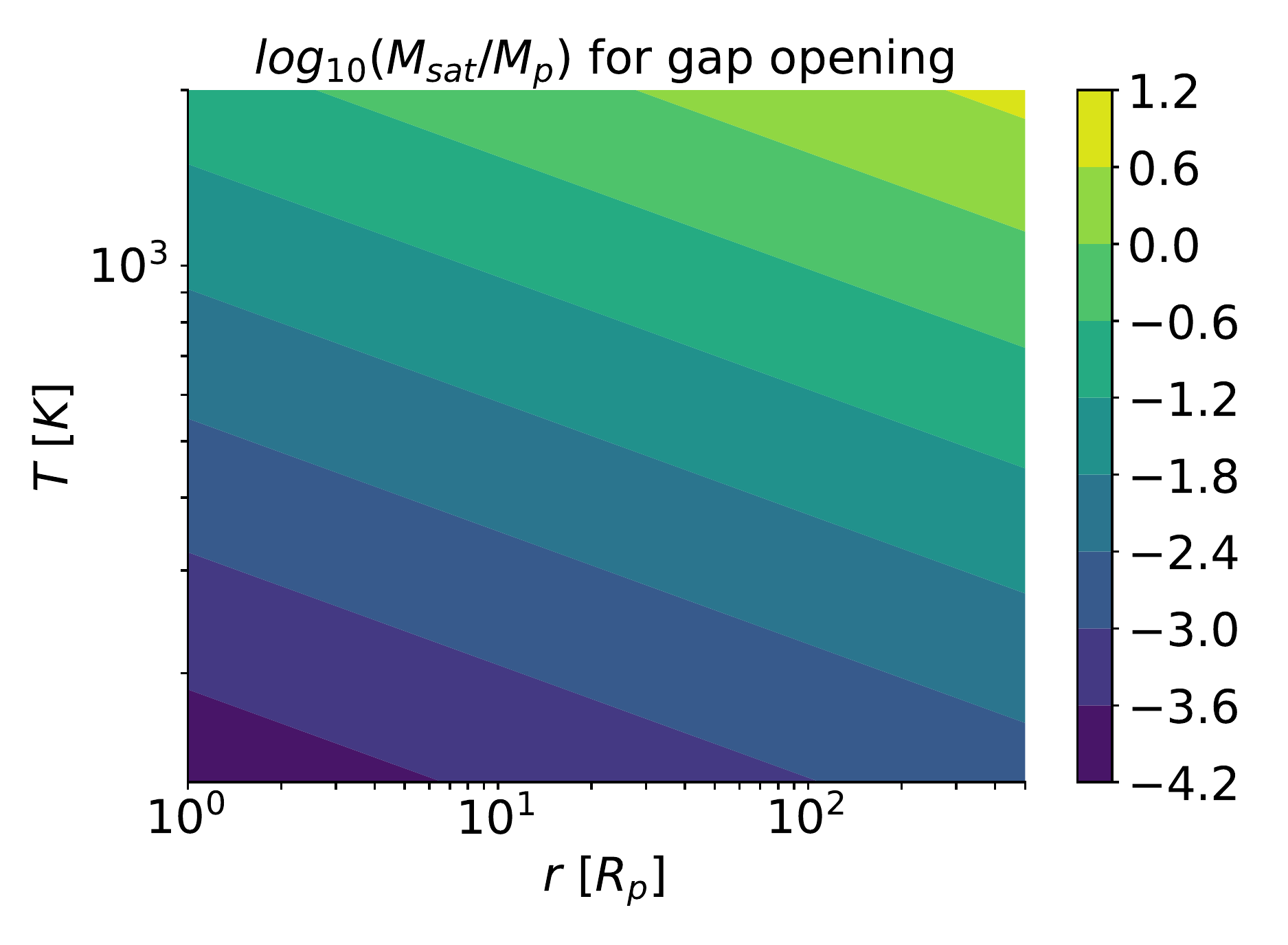}
\caption{Threshold mass for gap opening ($P<1$) as a function of $T$ and $r$. In the best configuration, i.e. low temperature close to the planet, a quite a big satellite is still needed to open a gap.}
\label{fig:gap_opening}
\end{figure}

We also investigated whether moons can open a gap at all in our model. First of all, one can notice that parameter $P$ (see Section \ref{form_and_evo}) depends only on the mass of the satellite, the temperature of the CPD, and the position of the satellite in the disc. Hence, it is possible to compute the satellite mass $M_s$ that can open a gap, as an analytic function of $r$ and $T$. This way we found that in our model it is very difficult to open a gap at all (Figure \ref{fig:gap_opening}), as already stated by \citealt{Moraes17}. In the best case (low temperature close to the central planet) a satellite with $M_s\simeq 10^{-4}M_p$ is needed, which is a quite high value considering the masses of the Galilean satellites distribute between $10^{-5}$ and $10^{-4} M_p$.

\subsection{The number of survived satellites}\label{numbers_mass}

\begin{figure}
\centering
\includegraphics[width=\columnwidth]{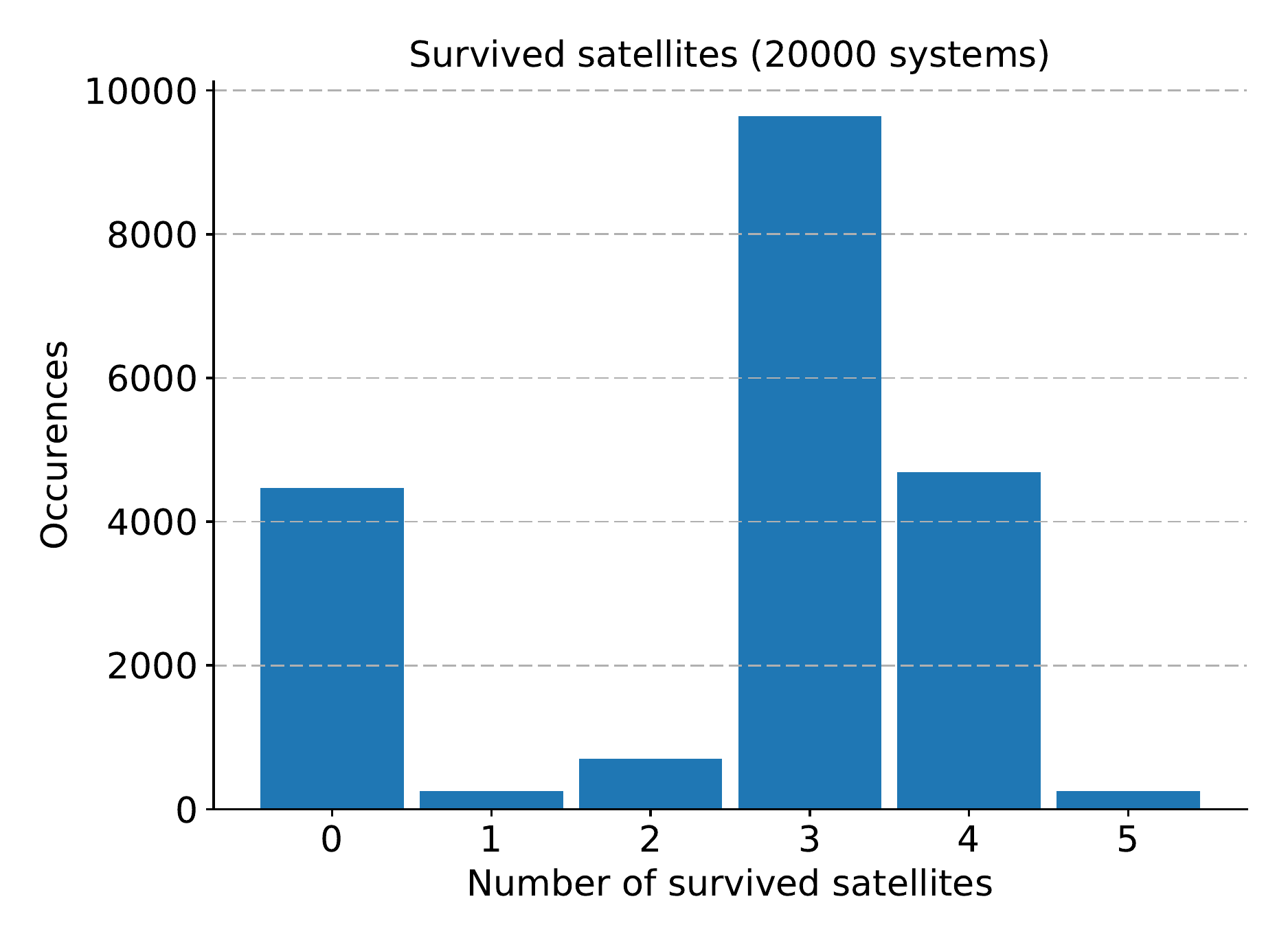}
\caption{Occurrences of systems with certain numbers of satellites. The peak of the distribution is at $3$, while the upper limit is at $5$. The most peculiar thing is the minimum visible between $1$ and $2$.}
\label{fig:histogram_numbers}
\end{figure}

In Figure \ref{fig:histogram_numbers} we show the satellites that prevail in each one of the 20000 systems after the gaseous CPD (and the PPD) dissipates.
In other words these are the moons that exist in the system when the gaseous CPD (and the PPD) dissipates. Without gas, the migration stops, therefore the dynamical evolution of the satellite system has been terminated. 
The histogram in Figure \ref{fig:histogram_numbers} shows that the most common outcome is a system with 3 satellites. The maximum number of satellites that can be formed in a system is 5.
While $4$ is the second most common result, no-survivor case is also frequent. The expectation is that the occurrence rate decreases with increasing amount of moons, however our results show an intriguing minimum at $N=1-2$.

To investigate the reason behind the minimum at 1-2 satellite masses, we used again the second type of population synthesis approach, varying separately the three initial parameters (as described in the first paragraph of Section \ref{results}): dust-to-gas ratio, $t_{\rm{disp}}$ and $t_{\rm{refilling}}$. We found that changing $t_{\rm{disp}}$ does not affect the distribution. This is because the migration timescale, which basically controls the number of coexisting (and then survived) satellites, does not depend on $t_{\rm{disp}}$. This confirms our considerations about $t_{LG}$ in Section \ref{t_LG}. While varying the dust-to-gas ratio, we arrived to the expected result: the more dust produces more satellites, hence more moons survive till the end of the evolution of the disc. 

\begin{figure} 
\centering
\includegraphics[width=\columnwidth]{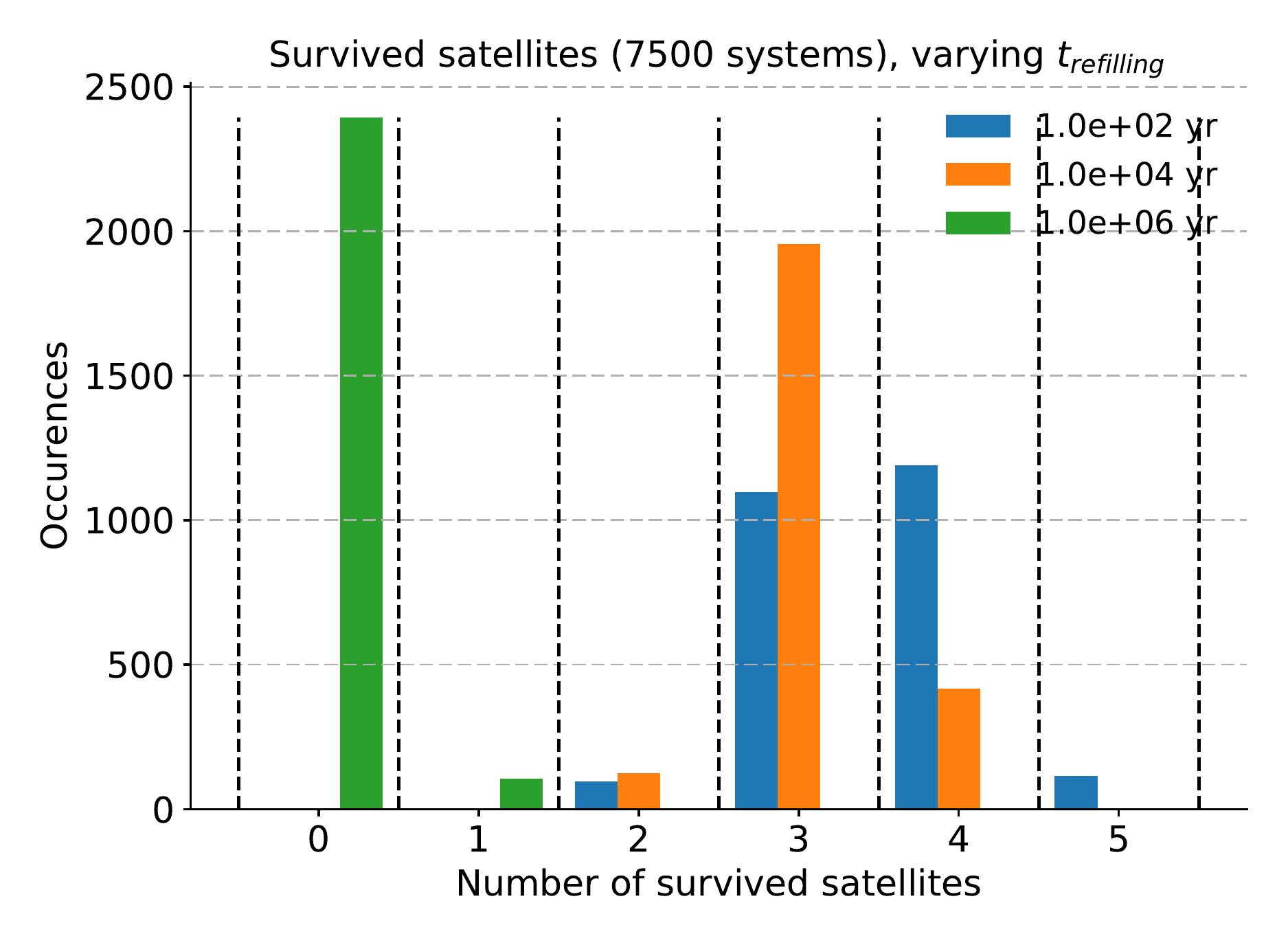}
\caption{The number of survived satellites when varying only $t_{\rm{refilling}}$.}
\label{fig:number_varying}
\end{figure}

The most extreme difference is found when the refilling mechanism timescale varies (Figure \ref{fig:number_varying}). In this case,
when refilling is slow, only 0-1 moons survive, while when refilling is fast, the distribution peaks at around 3, and this transition occurs between $t_{\rm{refilling}}=10^5$ and $t_{\rm{refilling}}=10^6$ years. The shape of the distribution does not change with varying this parameter, the minimum will be always at 2. With even narrower spacing in the transition region, we revealed that the transition is quite quick and it happens when $t_{\rm{refilling}}\simeq1-2 \times 10^5 yr$.

The reason behind this minimum at 2 is the following. In our model, as it was mentioned in Section \ref{form_and_evo}, embryo formation is triggered by two conditions: the dust trap has to be out of the previous satellite's feeding zone and the dust-to-gas ratio in the dust trap has to be $\ge 1$. The timescales of these two processes strongly depend on $t_{\rm{refilling}}$. The first one because migration depends on the satellites mass, therefore it also depends on the accretion rate. While, in turn, the accretion rate depends on the available dust and refilling, and the dust-to-gas ratio $\ge 1$ depends directly on $t_{\rm{refilling}}$. Given these two conditions, when refilling is fast enough to allow embryos to accrete and to move away from the feeding zone, it is already fast enough to allow dust-to-gas ratio in the dust trap to reach the value $1$ at least $3$ times in a migration timescale. This means that there would always be at least $3$ satellites at the same time, leading to usually $3$ survived moons at the end. It is possible to have a few systems with $1$ or $2$ satellites only when e.g. the disc evolution timescale is fast enough, as shown in Figure \ref{fig:number_varying}. 

\subsection{Formation temperature}\label{T_form}

The composition of the Galilean satellites are very diversified: while Io is completely rocky, the outer three contains some or significant amount of water \citep{Sohl02}. The water ice can be accreted to the body if it was formed in a disc below the water freezing point, therefore the water content of the Galilean satellites is a strong constraint that the temperature of the forming disc had to be below the water freezing point, $\sim 180 K$ \citep{Lodders03}. Therefore, we checked the temperature of the disc location where the last survived generation of satellites formed. Because in our model the satelletesimals form in the dust trap, and, most of the dusty material is also generated at this location, we defined a formed moon as icy if the dust trap temperature was below 180 K in our disc evolution, and as rocky if the temperature was higher than that. We found that 85 \% of survived satellites are icy, possibly coexisting with rocky ones. 

\begin{figure}
\centering
\includegraphics[width=\columnwidth]{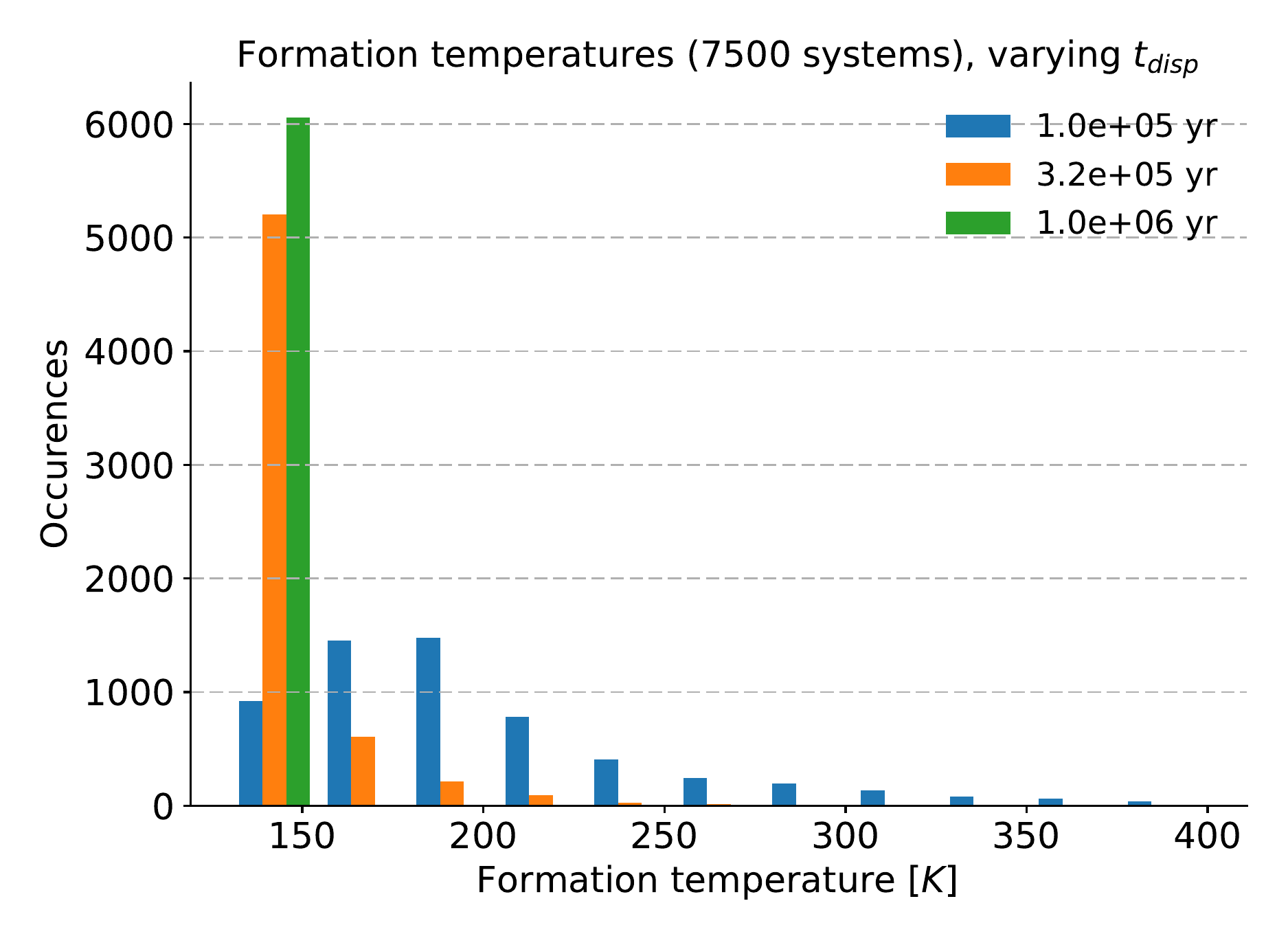}
\caption{The formation temperatures distribution changes when $t_{\rm{disp}}$ is varying. For shorter dispersion timescales the distribution is flatter, with a peak at about 200 K, while for longer dispersion timescales all the distribution concentrate around 130 K.}
\label{fig:T_form_varying}
\end{figure}

Like in the previous cases, we also checked how the formation temperature depends on the three parameters individually, which we varied in the population synthesis. The influence of both the dust-to-gas ratio and the $t_{\rm{refilling}}$ is trivial, because in this case they do not have any practical effect on temperatures and almost nothing changes when these two parameters vary (dust-to-gas ratio does not have effect because we considered an average dust-to-gas ration in computing temperature evolution in Section \ref{disc_evolution}). On the other hand, the disc dispersal timescale will affect the temperature evolution of the disc, through the opacities/optical depth. This is clearly visible in Figure \ref{fig:T_form_varying} where the formation temperature distribution changes shape and moving its peak from about $200$ K to $130$ K, as the dispersion timescale is longer.

\section{Discussion}\label{discussion}

As it is usual in population synthesis, the choices of the parameters, as well as some assumptions on the processes might change the results. In this Section we will discuss this, and describe tests and their results on the model, underlining also the biases that affect this work.

First of all, the disc structure has been modeled starting from the density and temperature profiles in the mid-plane of the disc coming from 3D radiative hydro simulations. All the other features of the disc, such as scale-height, pressure, surface density, sound speed, etc., have been computed from the 1D disc model \citep{Pringle81}. This is a first approximation that affects some of the CPD features, such as radial velocity profile, opacity and azimuthal velocity, since these quantities strongly depend on, for example, the pressure gradient in the mid-plane, that is computed from the 1D model. Furthermore, for this particular work we used a hydrodynamical model designed for CPD formation in core-accretion. If the CPD forms via disc instability, its properties would be significantly different, see e.g. \citet{Shabram13}, \citet{Szulagyi17a}.

Another bias is the disc evolution. For both dispersion and cooling we chose to use self-similar solutions, but, although modeling dispersion of the disc in this way is something common and already used in previous satellite population synthesis works \citep{Ida08, Miguel16}, a self-similar solution for cooling was a choice taken in order to be consistent with the rest of the semi-analytical framework, since it is the first time that CPD-cooling is performed in such a model. 

Whether or not there is a magnetospheric cavity between the planet and the disc can affect how many moons are lost in the planet, or whether they could be captured into resonance (easily). With no cavity between the planet and the disc, the migration rate of the moons will not be slowed down sufficiently and they will be easily lost in the planet. If there was a disc inner edge, that could hold the inner moons, and, behind, a resonance chain of satellites could pile up \citep{Sasaki10,Ogihara12}, like in the case of Super-Earths in PPDs \citep{Ogihara09}. Even in this case, the torque of the newly formed, outer satellites can eventually push the inner moon into the planet. Nevertheless, in this case probably less moons would be lost and more satellites in resonances would be the outcome. In the case of stars, due to the very strong magnetic fields, there is a gap between the surface of the star and the inner PPD. However, giant planets have significantly weaker magnetic fields, Jupiter, for example, has about 7 Gauss today \citep{Bolton17}. Even though it can be expected, like in the case of stars, that giant planets might have stronger magnetic field during their early years than today, no work has been carried out on this matter. There might be a scaling law between the luminosity and the magnetic field as it was pointed out by \citet{Christensen}, suggesting that a luminous planet in its formation phase could have a high magnetic field. On the other hand, \citet{OM16} calculates that Jupiter had to have at least an order of magnitude higher magnetic field than it has today, to induce magnetospheric accretion (and have a cavity between the planet and the disc), and the authors state that it is unlikely that Jupiter ever had such a strong magnetic field. They conclude, that the boundary layer accretion (i.e. when the disc touches the planet surface, like in our hydrodynamic simulations) is a more viable solution. Moreover, if the giant planet has a strong magnetic field, in itself this is not a sufficient condition for magnetospheric accretion to start. The gas inside the CPD has to be ionized, otherwise, the neutral gas will not care about the magnetic field and will enter into the cavity region. The ionization fraction of the CPD, on the contrary to the inner PPD, is very low as it was found in several works \citep{SzM17,Fujii11,Fujii14}.

Nevertheless, we checked how the results change when a cavity is assumed between the planet and the disc. In this case the first satellite would stop at the edge of the disc. The following satellite would then approach the first one and it would possibly be caught in a $2:1$ resonant configuration. Whether or not this capture happens can be inferred from analytical conditions, e.g. in \citet{Ogihara10}. In their work they found that, in case of a sharp disc edge and using the type I migration formula by \citet{Dangelo10} for its simplicity (we show below that changing the type I migration formula does not change our results significantly), up to 3 satellites would be locked in a resonant configuration when $t_e/t_a<1.7 \times 10^{-3}$, where $t_e$ is the eccentricity damping timescale and $t_a$ is the type I migration timescale. In our case this criterion implies a condition on the aspect ratio of the disc at the inner edge, i.e. $h/r < 0.024$. Using the definition of $h$ in a 1D disc model \citep{Pringle81} one finds the condition
\begin{equation}
\frac{T}{[K]}\frac{r_{\rm{cavity}}}{R_p} \le 210
\end{equation}
where $T$ is the temperature at the inner edge of the disc. This means that if we want to pile satellites up starting from the position of Io ($\simeq 6 R_p$) we need to have a temperature of about $35 K$, that is unphysical, due to the background temperature at Jupiter's location is about 130 K. 

Even if building a resonant structure is not possible in our model, we checked how the final results change when a cavity (as big as $2.5 R_p$ or $5R_p$) is introduced. In this case we considered that satellites stop their migration due to gas interaction when reaching the inner edge of the disc, but they still dynamically interact with other satellites. This means they still can be lost into the planet. The interaction between satellites has been modeled following the approach of \citet{Ida10}, i.e. considering that satellites tend to enlarge their orbital distance $\Delta a$ at each encounter. As expected, we found that we have more surviving satellites (their mean number grows from $2.5$ in the case without a cavity to $3.8$ and $4.5$ respectively, when the two different cavities are introduced) and as a consequence the integrated final moon mass grows from a median value of $6\times 10^{-4} M_p$ to $8\times 10^{-4} M_p$ and $12\times 10^{-4} M_p$, respectively, while the mean mass of single satellites does not change significantly (when satellites stop their migration in the cavity they also stop their accretion). Further investigations about surviving and lost satellites would need a more precise model for resonance capturing and for collisions between satellites, since this would be dominant processes in the satellite evolution.

In this work we also assumed that streaming instability forms the seeds of the moons. More conventional approaches based on collisional coagulation of dust grains would work with lower  dust-to-gas ratios, but would provide much longer formation timescales. In the latter models it is notoriously difficult to overcome the meter-size barrier as well as other issues. Since our hydrodynamical simulations have showed that dust traps appear in CPDs, it was natural to assume that streaming instability can operate. Another mechanism, that could have provided the seeds is the capturing of planetesimals from the PPD \citep{DP15,Tanigawa14}. Given that we found that the CPD is an efficient satelletesimal factory, we believe that there is no need for planetesimal capturing to form the moons there.

Regarding testing the initial parameters, we first checked the effect of initial embryo mass and a different Type I migration formula. In the latter, instead of the Paardekooper-formula \citep{Paardekooper11} we tested the $b_I$ coefficient from \citet{Dangelo10} and \citet{Dittkrist14}. Our finding is that the distribution of the population does not change much, the difference is within the change that is caused by random variations. 

In comparison to the previous satellite population synthesis work by \citet{Miguel16}, our results are somewhat different. While the other authors started with a Minimum Mass Sub-solar Nebula that is created by the current position and composition of Galilean moons, we use real hydrodynamic simulations on the circumplanetary disc as an initial gas and dust disc. Like them, we take into account the disc evolution both in dust and gas density, but we also account for evolution in the temperature profile, and we do not consider a cavity between the planet and the disc.  They found that in the case of long disc lifetimes, the survived satellites are less numerous and have lower masses than in our case, since the biggest ones have enough time to migrate and be lost into the central planet. The difference comes from the different dust-to-gas ratios, different disc initial parameters, and the assumption on which process generates the seeds of satellites, but also from the fact that they did not have any dust supply in the disc while accretion on protosatellites creates gaps in the dust profile. As a consequence their protosatellites have less available dust to grow to larger sizes.

Comparing also to the previous works by \citet{Canup02} and \cite{Canup06} we find that our results are partially in agreement with their conclusion, but we have also points of disagreement. First of all we agree that the CPD can in general be less massive than the MMSN model, in the scenario in which the disc itself is continuously fed by the influx from the PPD. We also agree on the fact that with our conditions on viscosity ($\alpha=4\times 10^{-3}$) type I migration should be always inward and on the fact that surviving satellites should form very late in the evolution of a Jupiter-like planet. On the other hand, we disagree on the fact that all the satellites should form, or have formed in the specific case of Jupiter, slowly, in $>10^5 yr$. 

The requirement of slow formation comes from the need to explain the non-differentiated nature of Callisto. However, in our model, since we can form many generations of satellites, up to the time when the disc is already more than a million years old, Callisto can form late and gradually. Furthermore, by starting with planetary cores formed by streaming instability, collisions between large planetary embryos, which would cause melting and differentiation, are not required to assemble the satellite.

\section{Conclusion}\label{conclusion}

In this work we investigated the formation and the evolution of the Galilean satellites in a circumplanetary disc around a Jupiter-like planet. We used a population synthesis approach involving 20000 systems, using the initial conditions (disc density and temperature profiles) from a 3D radiative simulation of \citet{Szulagyi17}, including the continuous feeding of gas and <mm sized dust from the protoplanetary disc \citep{Szulagyi14}. In the population synthesis, we accounted for the disc evolution and used a dust density profile from a realistic dust coagulation model of \citet{Drazkowska18}. Furthermore, in our model the seeds of the moons form via streaming instability in a dust trap, whose location is around 80 $R_{Jup}$ based on the vertical velocity profiles of the hydrodynamic simulation. The satellitesimals then migrate, accrete, are captured in resonances and are often lost in the planet. 

Nevertheless, we found that due to the dust trap, and the continuous influx of dust from the circumstellar disc, massive satellites are forming (the distribution peaks above the Galilean mass at  $\simeq 3\times10^{-4} M_J\simeq M_{Earth}$). Due to their high masses, they quickly migrate into the planet via Type I migration, because in most of the cases the gap opening criterion is not satisfied, the migration cannot enter the Type II regime. This means that the satellites form in sequence, and many are lost into the central planet polluting its envelope with metals. Our results show that the moons are forming fast, often within $10^4$ years (20 \% of the population), which is mainly due to the short orbital timescales of the circumplanetary disc.  Indeed the CPD completes several orders of magnitude more revolutions  around the planet than the protoplanetary disc material can do around the star at the location of Jupiter. Due to the short formation time, the satellites can form very late, about $30\%$ after 4 dispersion timescales, i.e. when the disc has $\sim 2\%$ of the initial mass.
Since our model included disc evolution, the CPD cooled off during this time, allowing to form icy moons, when the dust trap temperature dropped below 180 K, i.e. the water freezing point. We found out that about 85\% of the survived moons could contain water (ice). The production of moonlets and the migration rate provided such a situation, when the number of survived moons peaked around 3, but often no moons survived at all. 

The lost satellites bring on average 15 Earth-masses into the giant planet's envelope, polluting it with metals, that can contribute to the abundance of heavy elements in Jupiter's envelope. The high mass satellites we found in our population synthesis have intriguing implications for the future surveys of exomoons. Indeed even with the current instrumentation, an Earth-mass moon around a Jupiter analog can be detected if the planet is orbiting relatively close to its star \citep{Kipping09}. 

\section*{Acknowledgments}

We are thankful for anonymous referee for their comments that helped to clarify the paper. We also thank for the useful discussions with Yann Alibert, Christoph Mordasini, Clement Baruteau, Willy Kley and Ren\'e Heller. This work has been in part carried out within the frame of the National Centre for Competence in Research  ``PlanetS"  supported by  the  Swiss  National Science Foundation. J. Sz. acknowledges the support from the ETH Post-doctoral Fellowship from the Swiss Federal Institute of Technology (ETH Z\"urich) and the Swiss National Science Foundation (SNSF) Ambizione grant PZ00P2\_174115. Computations have been done on the ``M\"onch" machine hosted at the Swiss National Computational Centre.





\label{lastpage}
\end{document}